\documentclass[useAMS,usenatbib]{mn2e}
\usepackage{journals}
\usepackage[pdftex]{graphicx,color}
\usepackage[latin1]{inputenc}
\usepackage[T1]{fontenc}
\usepackage{graphics}
\usepackage{amsfonts}
\usepackage{amsmath}
\usepackage{multicol}
\usepackage{layout}
\usepackage{amssymb}
\usepackage{natbib}
\usepackage{color}
\usepackage[english]{babel}
\pdfoutput=1

\title[Simulated Strong Lenses]{Characterising Strong Lensing Galaxy Clusters \\
using the Millennium-XXL and MOKA simulations}
\author[Giocoli C. et al. 2016]{\parbox{\textwidth}{
    Carlo Giocoli$^1$\thanks{E-mail: carlo.giocoli@lam.fr}, 
    Mario Bonamigo$^1$, Marceau Limousin$^1$, Massimo Meneghetti$^{2,3}$,
    Lauro Moscardini$^{4,2,3}$, Raul E. Angulo$^5$, Giulia Despali$^{1,6}$, Eric Jullo$^1$}\\ $ $ \\
  $^1$ Aix Marseille  Universit\'e,  CNRS,  LAM (Laboratoire  d'Astrophysique  de Marseille)  UMR 7326,  13388 Marseille,  France \\
  $^2$ INAF - Osservatorio Astronomico di Bologna, via Ranzani 1, 40127 Bologna, Italy \\ 
  $^3$ INFN - Sezione di Bologna, viale Berti Pichat 6/2,  40127 Bologna, Italy \\
  $^4$ Dipartimento di Fisica e Astronomia, Alma Mater Studiorum Universit\`{a} di Bologna, viale Berti Pichat, 6/2, 40127 Bologna, Italy \\
  $^5$  Centro de Estudios de F\'isica del Cosmos de Arag\'on (CEFCA), Plaza San Juan 1, Planta-2,  44001 Teruel,  Spain\\
  $^6$ Max Planck Institute for Astrophysics, Karl-Schwarzschild-Strasse 1, 85740 Garching, Germany
}
\pubyear{2016}

\begin{document}

\maketitle

\begin{abstract}
In this paper  we investigate the strong lensing  statistics in galaxy
clusters.   We  extract dark  matter  haloes  from the  Millennium-XXL
simulation,  compute their  Einstein radius  distribution, and  find a
very good  agreement with  Monte Carlo  predictions produced  with the
MOKA code. The distribution of the Einstein radii is well described by
a log-normal distribution, with a considerable fraction of the largest
systems  boosted  by different  projection  effects.   We discuss  the
importance of substructures and triaxiality in shaping the size of the
critical lines for  cluster size haloes.  We then  model and interpret
the  different deviations,  accounting for  the presence  of a  Bright
Central  Galaxy   (BCG)  and   two  different  stellar   mass  density
profiles. We present scaling relations between weak lensing quantities
and the size of the Einstein radii. Finally we discuss how sensible is
the distribution of the Einstein  radii on the cosmological parameters
$\Omega_M - \sigma_8$ finding  that cosmologies with higher $\Omega_M$
and $\sigma_8$ possess a large  sample of strong lensing clusters. The
Einstein radius  distribution may help distinguish  Planck13 and WMAP7
cosmology at $3\sigma$.
\end{abstract}


\section{Introduction}

Spectroscopic   galaxy  redshift   surveys   and  numerical   $N$-body
simulations have revealed a large-scale  distribution of matter in the
Universe  featuring a  complex network  of interconnected  filamentary
galaxy                                                    associations
\citep{tormen04,springel05b,des05,sousbie08,sousbie11,guzzo14,percival14,lefevre15,codis15}.
Vertices, i.e.   interconnections among  the filaments,  correspond to
the very dense compact nodes within this \textit{cosmic web} where one
can           find          massive           galaxy          clusters
\citep{tormen98a,bryan98,shaw06,borgani11,bellagamba11}.

The  mass  density distribution  in  clusters  can be  inferred  using
different                    wavelength                   observations
\citep{meneghetti10b,donnarumma11,donahue16}.  In  particular, optical
and near-infrared  data provided by, for instance,  
the Subaru and the  Hubble Space
telescopes (HST)   are allowing  to indirectly
infer  the total  projected matter  density distribution  in clusters
through its  effect of gravitationally  bending the light of
background   galaxies   \citep{jullo07,merten15,limousin15}.
Gravitational   lensing,  as   predicted  by  the Einstein's   General
Relativity, deflects  light  rays once  they get  close to  a deep
potential well  \citep{einstein18,landau71}.  Light-rays  from distant
galaxies travelling in the space-time of our Universe can  
be  weakly  or  strongly  bent when they approach  a  galaxy  cluster 
\citep{bartelmann01,bartelmann10}.   The weak  lensing regime  happens
when the  light-rays travel far  from the  centre of the  cluster.  In
this case, the shapes of  background galaxies are only slightly altered
and, for a good determination of  the signal, it is usually necessary to
average    over  a    large    sample    of     background    systems
\citep{hoekstra12,hoekstra13,giocoli14,radovich15,formicola16}.    The
strong lensing regime takes place when the light-rays transit close to
the centre of the cluster, and the mass density becomes critical: the
lensing  event in  this case  is non-linear  and images  of background
galaxies  may  be  multiplied  and/or appear  stretched  and  elongated.
Depending on  the quality of the  data and on their  availability, weak
and strong lensing  data can be used separately or jointly   for a
better reconstruction  of the  projected mass from  the very central region  to the
outskirts of  the cluster.  In  the following, we will  concentrate
on the strong lensing regime and on the objects that originate it, which we will refer to as Strong Lensing Clusters (SLCs).

SLCs   may  constitute   a  peculiar   class  of
objects. While  their existence  is a  natural consequence  of General
Relativity,  ``giant arcs''  -- extremely distorted images of background galaxies -- hosted in  clusters have  been discovered
only  $30$ years  ago in  the core  of Abell  $370$, independently  by
\citet{lynds86}   and   \citet{soucail87}.    This   observation   was
recognised   by   \citet{paczynski87}   as  the   result   of   strong
gravitational lensing, a hypothesis later confirmed
by the measurement of the redshift of the arc \citep{soucail88a,soucail88b}.

Since then, SLCs have led to  many important 
advances in cosmology: ($i$) being a
direct  and  precise  probe  of  the  two-dimensional  projected  mass
density,  Strong  Lensing  (SL)   has  provided  accurate  mass  maps,
constraining  structure formation  properties and  evolution scenarios
\citep[for example:][]{broadhurst00,sand02,saha06,bradac06,zitrin09a,zitrin09b,newman11,verdugo11,sharon14};
($ii$) producing a natural gravitational amplification, SL has allowed
to    push    the    frontier    of    our    telescopes    \citep[for
  example:][]{richard06,coe13,atek14,zitrin14};  ($iii$)  providing a method to probe 
  the dark energy  equation of state, since
  images position depends on
the underling cosmology \citep[for example:][]{soucail04,jullo10}.

SLCs are now well established as a promising class of objects that cannot
be ignored in  cosmology, and their future is  extremely promising, since
future  facilities are  expected  to detect  thousands  of SLCs
\citep{euclidredbook,boldrin12,boldrin16,serjeant14},      and     the
exquisite resolution of the \textit{James Webb Space Telescope} (JWST)
will deliver unique multi-colour data sets for some of them. The growing
importance of SLCs has been recently illustrated by the CLASH  program
\citep{postman12} which has  been  awarded  of $500$  HST
orbits to observe $25$ massive SLCs.  More recently,
the  Hubble  Deep  Fields  Initiative has  unanimously  recommended  a
``Frontier  Field'' program  of  six deep  fields  concentrated on  SL
clusters  (together  with six  deep ``blank  fields'') in  order to
advance our  knowledge of the  early epochs of galaxy  formation and to
eventually     offer     a      glimpse     of     JWST's     universe
(http://www.stsci.edu/hst/campaigns/frontier-fields).    Each  cluster
will be  imaged with $140$ orbits,  leading to a total  of $840$ orbits
dedicated to the Frontier Field Initiative.

Very encouraging is also the  work performed by  \citet{zitrin11c}
on reconstructing the mass density distribution and the Einstein radius 
 (which estimates the size of the SL region) of a large sample
 of SDSS clusters. In  this  case, the  ``blind'
approach based on the assumption that light traces mass has allowed to establish that the Einstein radius distribution of clusters with $0.1 < z_l
< 0.55$ has a log-normal shape. Furthermore, a visual
inspection has revealed
that approximately $20$ percent of SLCs are boosted
by various projection effects. 

Given the significance of SLCs,  characterising this peculiar class of
object  is  crucial and  this  has  been  the  focus of  many  studies
\citep[for
  example:][]{hennawi07,meneghetti10a,redlich12,waizmann12}. This is also
the motivation  of the  present work, where  we aim  at characterising 
which clusters do generate strong lensing  features. Our approach is twofold:
($i$)  first  we will  use  the  large  sample of  cluster  statistics
afforded by  the Millennium -XXL simulation  \citep{angulo12} -- exploiting 
its large size (3 Gpc/$h$ box side), that allows to follows the formation of
many massive haloes;  ($ii$) second we will  complement the statistics
with a cosmological study based on clusters modelled using the MOKA code \citep{giocoli12a}.

We want to spend few words about the fact that the Einstein radius of lenses 
is not a direct observable quantity. The Einstein radius, defined by the location
of the tangential critical lines (more will be discussed about this in the first 
section) is a byproduct of the mass reconstruction pipeline by mean of 
parametric algorithms that typically assume that mass traces the light \citep{jullo07,zitrin11a} or
adaptively reconstruct the mass density distribution using non-parametric 
approaches \citep{merten14}.    

The paper is organised as follows: in Section~\ref{smethod} we present
the numerical  simulations and the pseudo-analytical  methods we adopt
as  bases for  our  analyses; in  Section~\ref{sslsr}  we discuss  the
scaling relations  between the  size of the  Einstein radius  and weak
lensing-derived quantities; in Section~\ref{scerd}  we present how the
Einstein  radius distribution  depends on  the matter  content of  the
universe  and on  the  initial normalisation  of  the power  spectrum.
Finally  in  Section~\ref{ssummary}  we   summarise  and  discuss  our
results.

\section{Methods}
In this paper we aim at  studying the strong lensing properties of galaxy
clusters --  through the  size of their  Einstein radius  -- extracted
from  a  very  large  cosmological box.  However,  the  limitation  of
possessing  the simulation only for 
one  cosmological  model  in
addition  to the  fact  that the  run has  been  performed only  using
collisionless  dark matter  particles  forced us  to complement  the
analyses using a pseudo-analytic approach to simulate convergence maps
of triaxial clusters. This latter method allows us, in a more flexible
way,  to investigate  which properties of clusters mainly  contribute in
shaping  the Einstein  radius,  to quantify  the  contribution of  the
stellar  component   and  to   understand  how  the   Einstein  radius
distribution  of   clusters  may   depend  on   specific  cosmological
parameters.

\label{smethod}
\subsection{Strong lensing of Clusters in the Millennium-XXL Simulation}
With a  box side of 3  Gpc$/h$, the Millennium-XXL  (M-XXL) simulation
\citep{angulo12} was especially tailored to study massive haloes which
can be  only found in very  large volumes, because of  their nature of
extremely rare objects.  The $6720^3 \sim 3\times 10^{11}$ dark matter
particles  have   a  mass  of  $6.174\times   10^9  M_{\odot}/h$;  the
Plummer-equivalent  softening length  is $\epsilon  = 13.7$ kpc.  For
reasons   of   consistency   with   the   previous   Millennium   runs
\citep{springel05b,boylan-kolchin09},  the   adopted  $\mathrm{\Lambda
  CDM}$ cosmology  as the following parameters total matter density  $\Omega_M= 0.25$, baryons
density  $\Omega_b=0.045$,  cosmological  constant  $\Omega_{\Lambda}=
0.75$, power spectrum normalisation  $\sigma_8= 0.9$ and dimensionless
Hubble parameter in $H_0/100\;\mathrm{km}/s/\mathrm{Mpc}$  $h=0.73$. We  remind the  reader that  the simulated
volume of the M-XXL is equivalent  to the whole observable Universe up
to redshift $z=0.72$.

\begin{figure*}
\includegraphics[width=\textwidth]{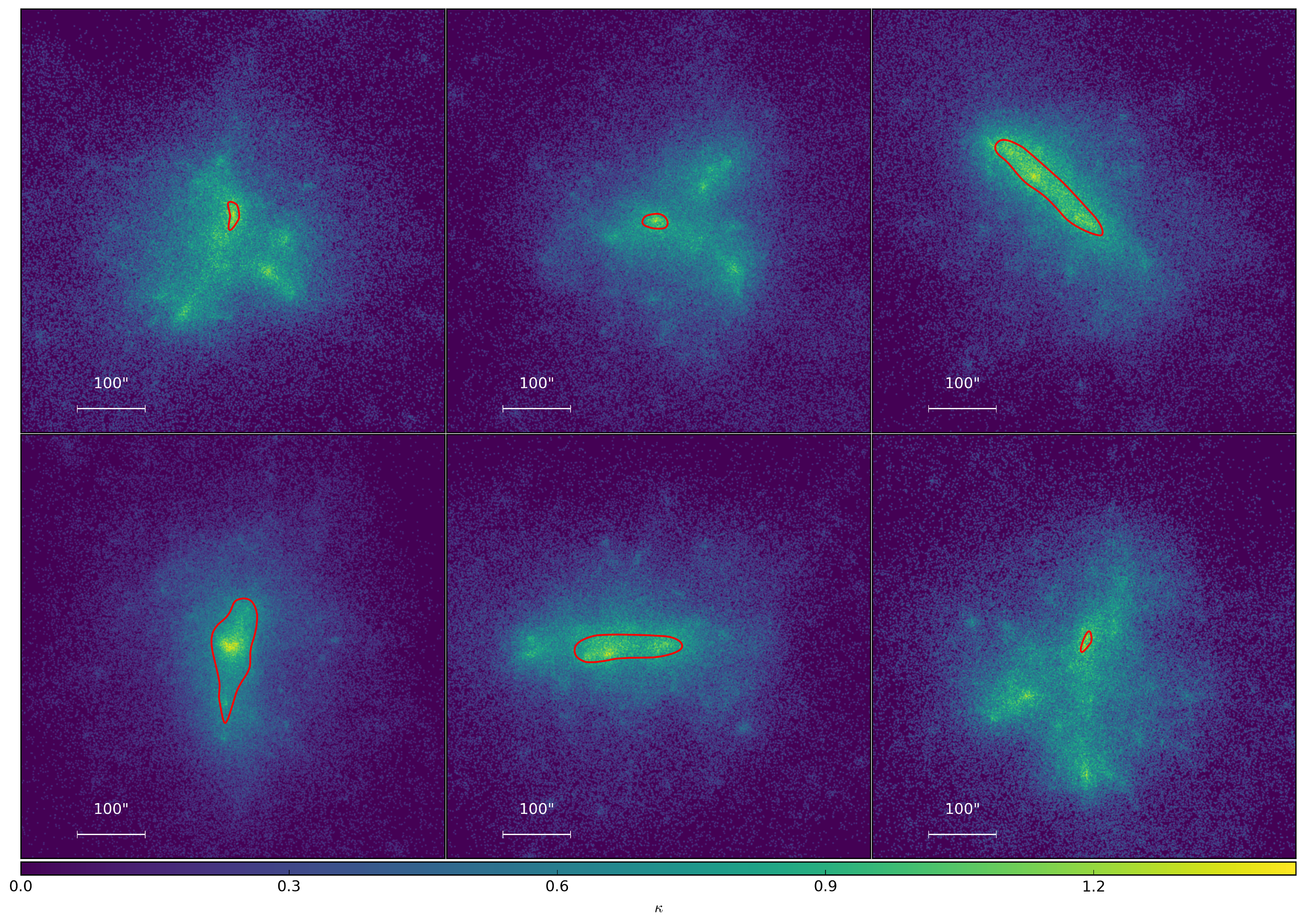}
\caption{Convergence  maps   of   different
  projections of  a halo extracted from  the Millennium-XXL simulation
  with  mass   $M_{200}  =  1.2  \times   10^{15}M_{\odot}/h$. 
  The red curves in each panel represent the tangential
  critical lines  from which we  compute the median Einstein  radii.   
  The top-three  images show  the three projections along the cartesian
  axes (i.e. \textit{random} with respect to the cluster morphology),
  while the bottom ones from left to right, are the projections
  along the  major, intermediate  and minor axes,  respectively.  This
  particular cluster  has the  peculiarity of  having in  one projection
  (namely the one in the left bottom panel) the    largest     Einstein    radius    in    our sample: $75$ arcsec. \label{figMXXLCluster}}
\end{figure*}

At each simulation  snapshot, haloes have been identified  using a FoF
algorithm. For  each FoF-group, starting  from the particle  with the
minimum potential, we  then compute $M_{200}$ as the  mass enclosing a
sphere $200$ times  denser than the critical density  $\rho_c$ at that
redshift.  In  our analysis  -- for the  motivation we  will underline
later -- we will  consider the halo catalogue at $z=1$ and
the corresponding snapshot  files.  Due to the large  number of haloes
identified in the  simulation volume we restrict our  analysis only to
the ones more massive than  $3\times 10^{14} M_{\odot}/h$ -- corresponding to
3135 systems.  For each halo respecting this criterion we store all the
particles  enclosed  in a  cube  of  $8\,\mathrm{Mpc}/h$ by  side and
project them in  a 2D-mass map resolved with  $2048\times 2048$ pixels
using  the  Triangular  Shape  Cloud technique,  along  six  different
directions. In the first three  cases we consider three projections along
the cartesian axes, which are then \textit{random} with respect to the cluster
morphology, 
 we then consider  three \textit{peculiar
  projections}  i.e.    along  the  ellipsoid  axes   as  computed  in
\citet{bonamigo15}: major, intermediate and minor axes. In placing the
particles on the  grid, to avoid particle  noise effects \citep{rau13,angulo14}
due to the  discreteness of the dark matter density, we apply a
Gaussian filter  with a scale  of $3.25$ kpc/$h$, which  corresponds to
approximately   one  third   of   the  simulation   Plummer-equivalent
softening.

\begin{figure*}
  \centering
  \includegraphics[width=0.45\hsize]{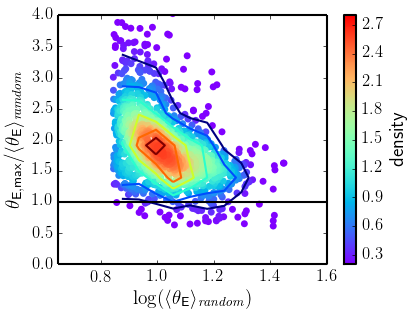}
  \includegraphics[width=0.45\hsize]{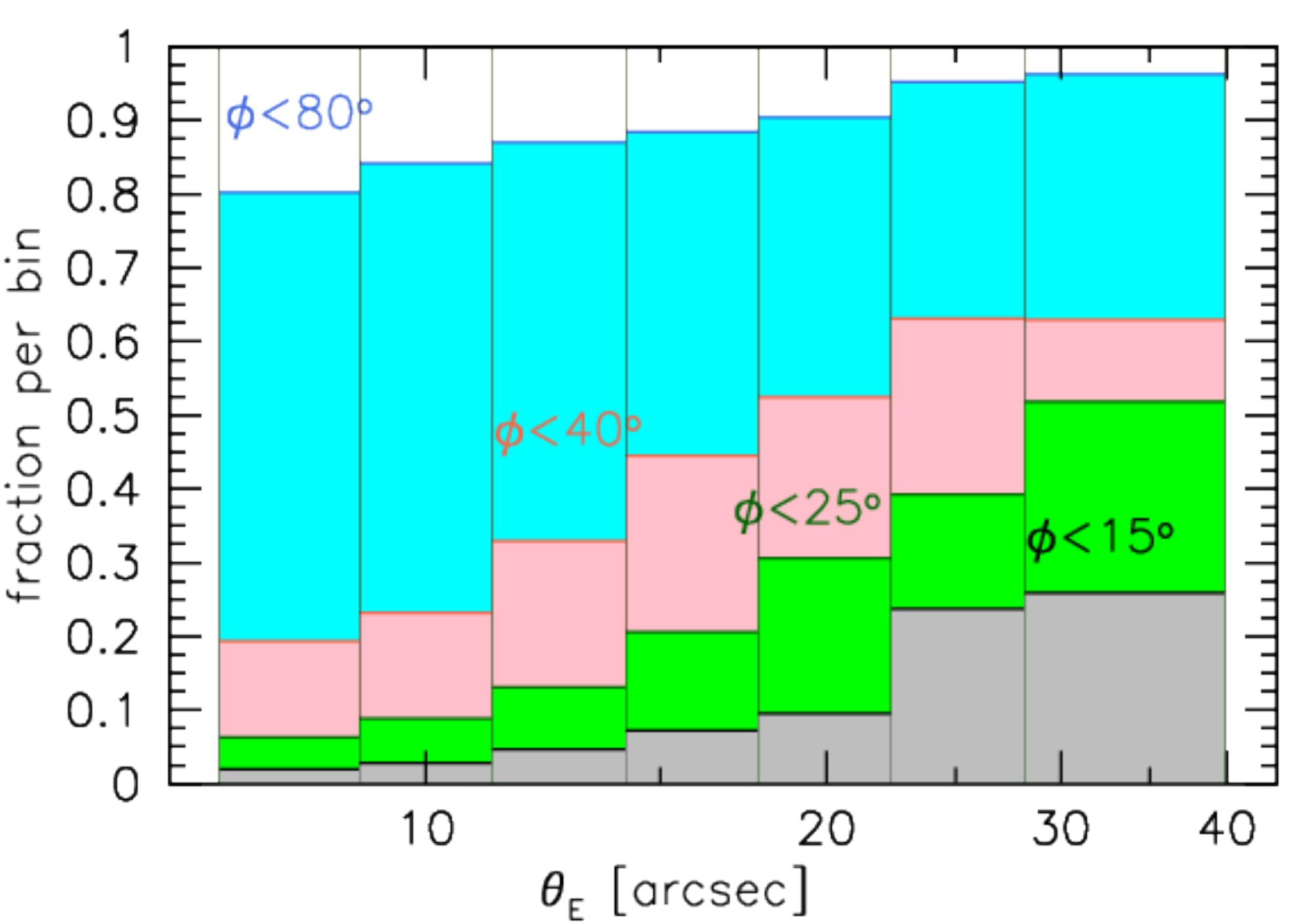}  
  \caption{Left  panel:  scatter plots  of  the  relative size  of  the
    Einstein radii  when the cluster  major axis of the  ellipsoid is
    oriented along the line-of-sight with compared to the average value
    of the three  \textit{random} projections: $\langle
    \theta_E  \rangle_{\it random}$.  Right  panel:  Fraction  i  of clusters  with  an  angle $\phi$  between  the
    direction  of  the  major  axis  of the  mass  ellipsoid  and  the
    line-of-sight smaller than  $80^\circ$, $40^\circ$, $25^\circ$ and
    $10^\circ$ as a function of the Einstein radius.
    \label{figtmaxtxyz}}
\end{figure*}

From  the constructed  mass  density maps  $\Sigma(x_1,x_2)$ --  where
$x_1$ and $x_2$ are the two cartesian coordinates on the 2D map projected in the
plane of the sky -- we compute the convergence $\kappa(x_1,x_2)$ as:
\begin{equation}
\kappa (x_1,x_2) = \dfrac{\Sigma(x_1,x_2)}{\Sigma_{\rm crit}}
\end{equation}
with 
\begin{equation}
\Sigma_{\rm crit} \equiv \dfrac{c^2}{4 \pi G} \dfrac{D_l}{D_s D_{ls}}
\equiv \dfrac{c^2}{4 \pi G}\dfrac{1}{D_{lens}}
\end{equation}
where  $c$  represents  the  speed  of light  and  $G$  the  universal
gravitational  constant; $D_l$,  $D_s$  and $D_{ls}$  are the  angular
diameter   distances   between  observer-lens,   observer-source   and
source-lens,  respectively;  we  also   define  the  lensing  distance
$D_{lens} \equiv D_{ls}D_s/D_l$.  We assume  clusters to be located at
$z_l=0.5$ and  sources at $z_s=2.5$, computing  the distances assuming
the  cosmological parameters  in agreement  with the  Planck13 results
\citep{planck1_14}: the matter density parameter $\Omega_M=0.307$, the
contribution  of  $\Lambda$ $\Omega_{\Lambda}=0.693$,  the  normalised
Hubble constant $h=0.6777$ and the  normalisation of the initial power
spectrum  $\sigma_8=0.829$.   We do  so  because,  even if  the  M-XXL
simulation  has  been  run  with   a  different  set  of  cosmological
parameters, we  assume to be able  to rescale those clusters  at $z=1$
from a M-XXL cosmology to a sample at $z=0.5$ in a Planck13 cosmology.
This is supported by the fact that the halo properties at $z=1$ in the
M-XXL cosmology  are very similar  to those  at $z=0.5$ in  a Planck13
cosmology  \citep{sheth99b,maccio08,zhao09,giocoli12b,despali15}; even
if  the two  mass functions  for haloes  more massive  than $3  \times
10^{14}M_{\odot}/h$  may be  different by  more than  $50\%$, the  two
concentration-mass relations deviate by less  than $5\%$.  \\ From the
convergence we can define the effective lensing potential as:
\begin{equation}
 \Phi(x_1,x_2) \equiv \dfrac{1}{\pi} \int \kappa(\mathbf{x}') \ln | \mathbf{x} - \mathbf{x}' | \mathrm{d}^2 \mathbf{x}',
\end{equation}
with $\textbf{x} \equiv (x_1,x_2)$, and then the pseudo-vector field of the shear $\mathbf{\gamma}=\gamma_1 + i \gamma_2$
as:
\begin{eqnarray}
 \gamma_1 (x_1,x_2) &=& \dfrac{1}{2}\left( \Phi_{11} - \Phi_{22}\right), \\ 
 \gamma_2 (x_1,x_2) &=& \Phi_{12} = \Phi_{21}\;
\end{eqnarray}
with $\Phi_{ij}$ representing the $i$ and $j$ derivatives of the effective lensing potential \citep{bartelmann01,bacon10}.
At first order, gravitational lensing induces distortion and stretch 
on background sources: typically a circular source is mapped through gravitational lensing
into an ellipse when both $k$ and $\gamma$ are different from zero. These effects 
are described by the Jacobian matrix:
\begin{equation}
A = 
\left( \begin{array}{cc}
1-\kappa-\gamma_1  &  -\gamma_2 \\
-\gamma_2 &  1-\kappa+\gamma_1 \end{array} \right)
\,.
\end{equation} 
The magnification is quantified as the inverse determinant of the 
Jacobian matrix that can be read as:
\begin{equation}
\mu \equiv \dfrac{1}{\mathrm{det}A}  =  \dfrac{1}{(1-\kappa)^2 - \gamma^2};
\end{equation}
the inverse of the eigenvalues of the Jacobian matrix measure the amplification in radial and tangential direction of background sources:
\begin{eqnarray}
\mu_r &=& \dfrac{1}{1 - \kappa + \gamma} \\  
\mu_t &=& \dfrac{1}{1 - \kappa - \gamma}.
\end{eqnarray}
For circularly symmetric lenses, the regions  in the  image plane  where the  denominator of  the relations above is equal to zero define  where the source images are infinitely radially and tangentially magnified, respectively.  In particular 
images forming close to the tangential critical curve are strongly 
distorted tangentially to it.
 
The definition of critical curves is more complex and 
 non trivial  in  asymmetric, substructured and triaxial clusters.
 From each  convergence map  the lensing potential  and the  shear are
 numerically  computed  in  Fourier  space  \footnote{using  the  FFTW
   libraries:  http://www.fftw.org} where  derivatives are  easily and
 efficiently calculated.   To avoid  artificial boundary  effects each
 map  is enclosed  in a  zero-padded  region of  $1024$ pixels. 
 We have tested the impact of the size of the zero-padded regions
 on the weak and strong lensing properties of individual non-periodic cluster maps
 and find that artefact mirror clusters do not appear when the size of the zero region 
 is at least half of the considered field of view. To define the Einstein radius of 
 the cluster we identify in the cluster maps points  of infinite tangential 
 magnification $\theta_t$ and define the
 Einstein radius  $\theta_E$ as  the median  distance  
 of  these points
 from the cluster centre:
 \begin{equation}
 \theta_E \equiv \mathrm{med} \left\{ \sqrt{\left(\theta_{i,x_1} - \theta_{c,x_1}\right)^2 +
 \left(\theta_{i,x_2} - \theta_{c,x_2}\right)^2 } \;\big{|}\; \theta_i \in \theta_t \right\}\,.
 \end{equation}
We   define the center of the cluster $\theta_c$ as the position of the particle  with  minimum  potential 
and the connected region defined by the tangential critical points $\theta_t$,
 when  they  exist, have to 
 enclose the cluster  centre; this ensures that  the critical points
 are not eventually assigned to a substructure present in the field of
 view. The robustness of this definition has already been tested and discussed 
 in a series of works \citep{meneghetti08,meneghetti10a,redlich12,giocoli14} to which we remind the reader for more details.  The size  of the  Einstein radius  defines a  measure of  the  strong  lensing region  and,  for an axially symmetric lens,
 permits to  estimate  the mass  enclosed within it using the equation:
\begin{equation}
 \theta_E = \left(\dfrac{4 G M(<\theta_E)}{c^2} \dfrac{D_{ls}}{D_l D_s}\right)^{1/2}
\end{equation}
assuming   that  all   mass  is   located   at  the   centre  of   the
lens. By geometrically measuring the area  $A$ enclosed by the tangential
critical curve it is possible  to define the effective Einstein radius
as  $\theta_{E,eff} =  \sqrt{A/\pi}$.  However,  we will  rely on  the
median   Einstein   radius   definition   that  --   as   noticed   by
\citet{meneghetti11}  and  \citet{giocoli14}  -- better  captures  the
presence of asymmetries  of the matter distribution  towards the cluster
centre.

In Figure~\ref{figMXXLCluster} we show  the six considered projections
of the halo  which in one them has the  largest Einstein radius ($75$ arcsec) in
our constructed  catalogue   -- namely in  the bottom
left panel.   The top  panels show  the $x$,  $y$ and  $z$ projections,
while the  bottom ones the  projections along the  major, intermediate
and minor axis of the halo ellipsoid, from left to right respectively.
In each  panel, the  red curves represent  the tangential  critical curves,
i.e.   where  images  of   background  galaxies  would  appear  highly
tangentially magnified  if located  close to the  optical axis  of the
lens  system.  From  the figure  we notice  that the  largest Einstein
radius occurs,  in this  case -- as in most of  the cases,  when the
major  axis   of  the   cluster  ellipsoid   is  oriented   along  the
line-of-sight; the  opposite holds when the minor  axis points towards
the observer.
\begin{table}
\centering
\caption{Percentage of the projections along which M-XXL clusters have
  the  largest  Einstein  radius   for  the  \textit{random}  and  the
  \textit{peculiar} projections, respectively.\label{tabthetaMXXL}}
\begin{tabular}{r|c}
  projection & $\%$ (\textit{random}) \\ \hline  $x$ & $34\%$ \\ $y$ &
  $32\%$   \\   $z$   &   $34\%$   \\   \hline   projection   &   $\%$
  (\textit{peculiar}) \\ \hline  major axis of the  ellipsoid & $86\%$
  \\ intermediate & $12\%$ \\ minor & $2\%$ \\ \hline
\end{tabular}
\end{table}
From the measured Einstein radius of each of the six projections of
all  clusters  in the  M-XXL  we  can summarise  (as it can  be read  in
Table~\ref{tabthetaMXXL}) that in  the \textit{random projections} the
probability of  having the largest  Einstein radius is uniform  in the
three  cases as expected. However,  considering  the \textit{peculiar  projections},
sample we  notice that  in $86\%$  of the  cases the  largest Einstein
radius appears when the major axis of the ellipsoid is oriented along the
line-of-sight  and in  $12\%$  ($2\%$)  of the  cases  when 
the orientation is the  intermediate (minor) axis.  We have
investigated those latter cases and they arise either ($i$) when there
is  a merging  event  which manifests  in the  presence  of a  massive
substructure projected in correspondence  of the cluster centre and/or
($ii$) when  the cluster ellipsoid is  very elongated in the  plane of
the sky.

In the left panel of  Figure~\ref{figtmaxtxyz} we quantify by how much
the Einstein radius grows when the cluster is oriented along the major
axis of its mass ellipsoid. We consider all clusters having at least
an  Einstein radius  of $7$  arcsec along one  of the  considered
projections\footnote{The  value of  $7$  arcsec  ensures that  the
  measurement of the size of the Einstein radius of the cluster is not
  affected nor  by particle noise neither  by the finite grid  size of
  the map.}.  In  this case we compare the size  of the Einstein radius
computed  when the  cluster  is  oriented along  the  major axis  with
respect to  the average value  measured from its  three \textit{random
  projections}.  From the  figure we observe that the  typical size of
an  Einstein radius  may grow  up to  a factor  of two/three  when the
cluster is  aligned along the line  of sight with respect  to a random
orientation;  we  also notice  some  cases  where the  Einstein  radius
computed  in a  \textit{random projection}  is larger  than the  value
measured when the mass ellipsoid  is oriented along the line-of-sight;
as  discussed previously  we verified that those  cases are  merging  clusters or  very
elongated ellipsoids  in the plane of  the sky.  All this  brings more
light to the general picture that  most of the strong lensing clusters
may possess their dark matter  halo major axis preferentially pointing
close to the line-of-sight \citep{oguri09a}. This is more evident in the right panel of
the same figure where we show  the fraction of SLCs per different bins
in $\theta_E$  that
possess an angle $\phi$  between the major
axis  of the  ellipsoid and  the  line-of-sight smaller than a given value: $65\%$  of SLCs  with
$30<\theta_E<40$ have an  angle $\phi$ between the  direction of their
major axis and the line-of-sight smaller than $40$ degrees.  Our finding are 
quite consistent with the results presented by \citet{oguri05} where the authors also discuss that 
the apparent steep observed mass profile can be reconciled with theoretical models if the triaxial ellipsoid of the 
dark matter halo is preferentially oriented with the major axis along the line-of-sight.

\begin{figure}
  \centering
  \includegraphics[width=\columnwidth]{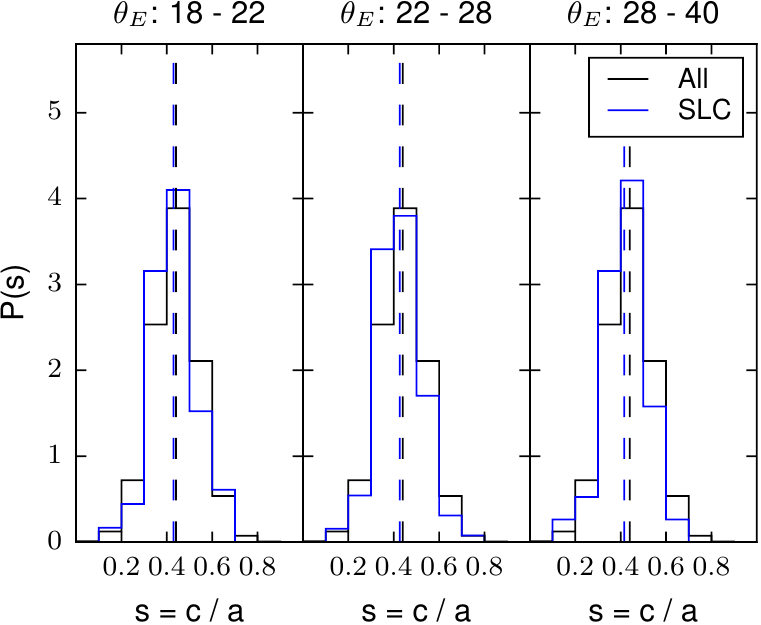}
  \caption{Probability distribution  functions of  the minor  to major
    axis ratio of the overall  M-XXL cluster population (black histogram) and
    of the  SLCs (blue histogram),  with each panel representing  a different
    bin  in $\theta_E$.  Vertical dashed  lines indicate  the mean  of
    the corresponding sample.
    \label{figpdfshape}
    }
\end{figure}
However, when looking  at random projections in the plane  of the sky,
the   sole   effect   of   triaxiality  is   less   obvious.    Figure
\ref{figpdfshape}  shows  the  difference  that  might  arise  in  the
distribution of shapes  -- namely minor to major axis  ratio $s$ -- by
selecting clusters that are strong lenses (blue histograms) instead of
the general population (black histograms). Haloes have been subdivided
in  bins of  Einstein radius  $\theta_E$,  each shown  in a  different
panel.   Even though,  as previously  found by  \citet{hennawi07}, the
distribution of  the axis ratio of  SLCs does not seem  to differ from
the distribution of the overall  population, the mean values (vertical
dashed  lines)  vary  up  to  $5\%$,  in  particular  for  very  large
$\theta_E$. A  Kolmogorov-Smirnov test showed  that we can  reject the
hypothesis that  the samples are  taken from the same  distribution at
significance level of $10\%$, meaning  that there is a low probability
 that  SLC have  the same  shape properties  of the
overall population. This suggests that the  concentration 
is mainly responsible in driving the correlation of the cluster Einstein radii.  

It  is  important  to  underline  that  the  effect  of
correlated and uncorrelated large scale  structures may also impact 
the  lensing properties  of  galaxy clusters  and  boost their  strong
lensing  cross section  as well  as the  size of  the Einstein  radius
\citep{puchwein09}.  Usually  to quantify  the impact  of uncorrelated
structures along  the line-of-sight it  is necessary to  run expensive
multi-plane          ray-tracing          lensing          simulations
 of    clusters   and   matter
extracted from cosmological  runs \citep{hilbert08,petkova14,giocoli15b}, things that are  beyond the purpose
of this paper.  However the  effect of correlated  structures on
the  lensing properties  can  be studied  selecting  for each  cluster
projection a larger  region along the line-of-sight,  and quantify how
these changes on the determination of the Einstein radius.  To do so, we
have  produced  two other  sets  of  convergence maps,  one  selecting
particles from a region of $16$  Mpc$/h$ and another from $32$ Mpc$/h$
along the line-of-sight, and projecting all  of them into a single lens
plane. We still keep the size of the region in the plane of the sky to be
$8$ Mcp/$h$ of a side. 
\begin{figure}
\includegraphics[width=\columnwidth]{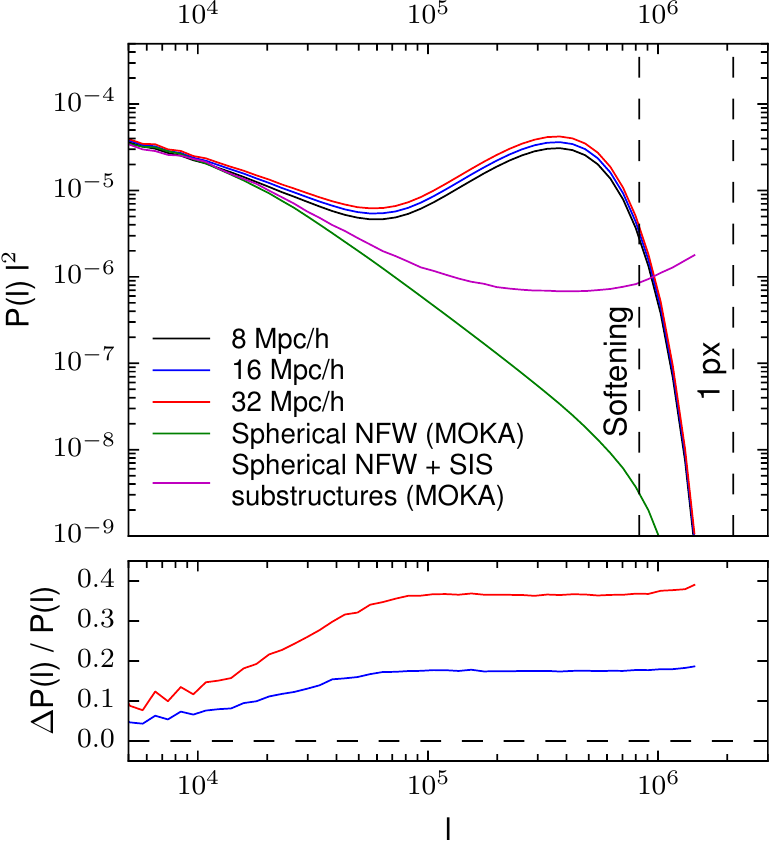}
\caption{Average convergence power spectrum  of haloes of
  the  M-XXL simulation.   Black, blue  and red  curve show  the average power
  spectrum derived extracting the particles contained in a region of $8$, $16$
  and $32$ Mpc$/h$ along the line-of-sight; in the plane of the sky in
  all three  cases we  have considered  particles in  a square  of $8$
  Mpc$/h$ of  side.   The green  curve shows  the prediction  from
  smooth  NFW  spherical  haloes,  while the  purple  one  presents  the
  prediction  for spherical MOKA  haloes with  substructures modelled
  with a Singular Isothermal Sphere (SIS) profile. The bottom panel shows the 
  relative residuals of the average power spectra measured using $16$ and $32$ Mpc/$h$ 
  with respect to the one computed assuming $8$ Mpc/$h$ as box side along the line-of-sight. \label{figpower}}
\end{figure}
As an example, in Figure~\ref{figpower} we show
the  average  convergence  power   spectra  of   the  \textit{random
  projections}  sample considering  a  region of  $8$,  $16$ and  $32$
Mpc/$h$ along the  line-of-sight in black, blue and red, respectively.  
In the bottom panel  we present the
relative  residuals of  the last  two cases  with respect  to the  $8$
Mpc/$h$ reference  one.  We notice  that the inclusion of  more matter
along  the  line-of-sight  tends  to increase  the  convergence  power
spectrum at small scales of  about $20$ percent for $16$ Mpc/$h$
and almost $40$ percent for $32$ Mpc/$h$, which however contains $4$
times the  volume.  In the figure  we show also for comparison the  
prediction from a
smooth NFW  halo (green curve) and  the power spectrum of  a spherical
halo (with the same large scale normalisation) with substructures
(in magenta): both  curves are obtained by averaging produced using MOKA haloes (see below), with
the same masses  and NFW concentrations of the M-XXL  sample.  In this
case we observe that the presence  of substructures in a halo tends to
increase the small scale power of more than one order of magnitude for
$l\gtrsim  3\times 10^4$  with respect  to a  smooth case.   The other
interesting  behaviour is  that  while the  power  spectrum of  haloes
extracted  from the  M-XXL are  characterised at  small scales  by the
particle noise and finite grid size of the maps \citep{vale03}, MOKA  
haloes  are particle  noise-free  and the  only
numerical limitation at small scale is set by the desired grid size of
the map.

\citet{puchwein09}  have  shown  that  the  presence  of  uncorrelated
structures tends  to boost both  the strong lensing  cross-section for
giant arcs  and the  size of  the Einstein  radii.  As  discussed, an
accurate description  of the contribution of  uncorrelated large-scale
structures needs expensive multi-plane  ray-tracing simulations and is
beyond the purpose of this paper.
\begin{figure*}
\includegraphics[width=0.49\hsize]{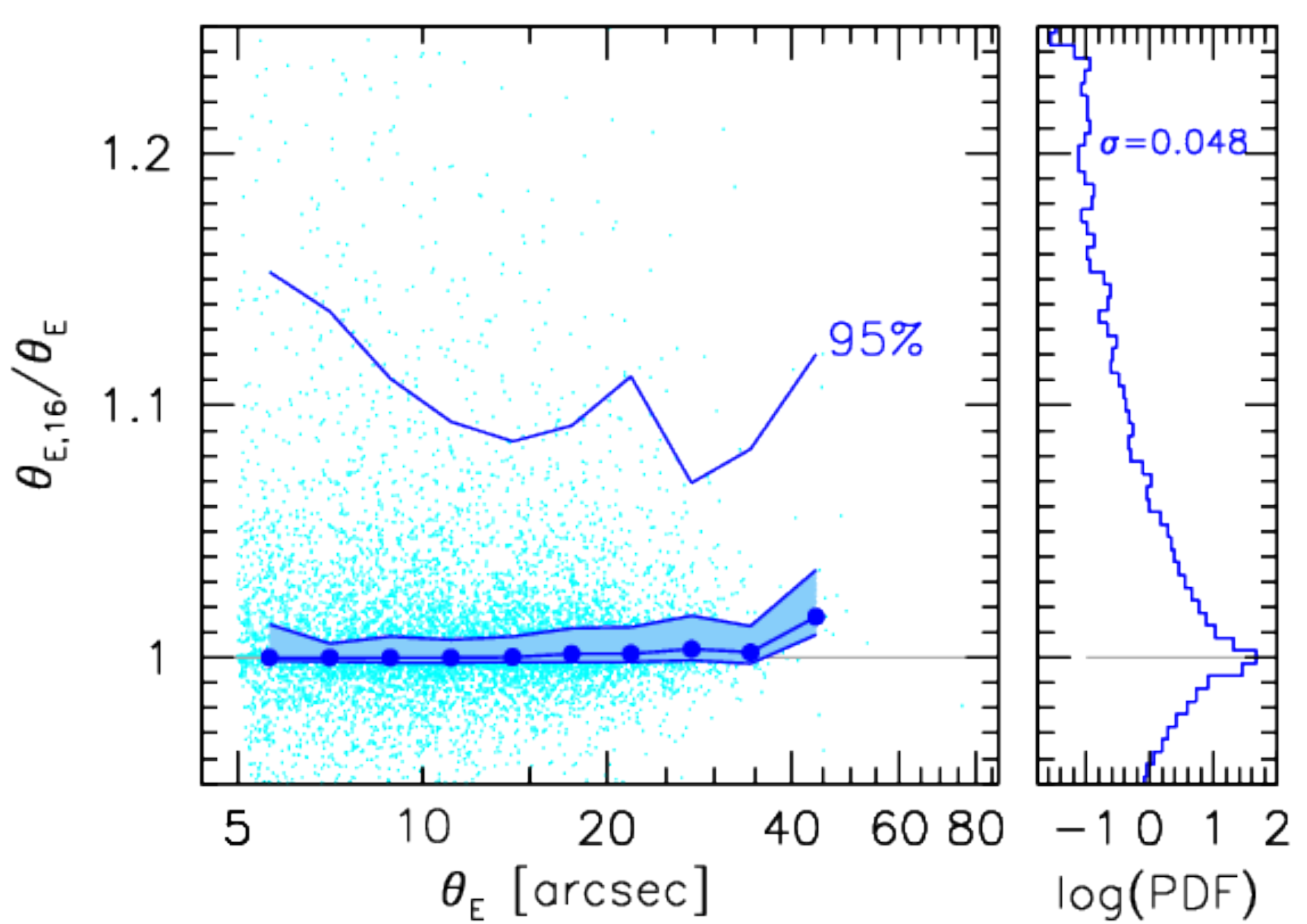}
\includegraphics[width=0.49\hsize]{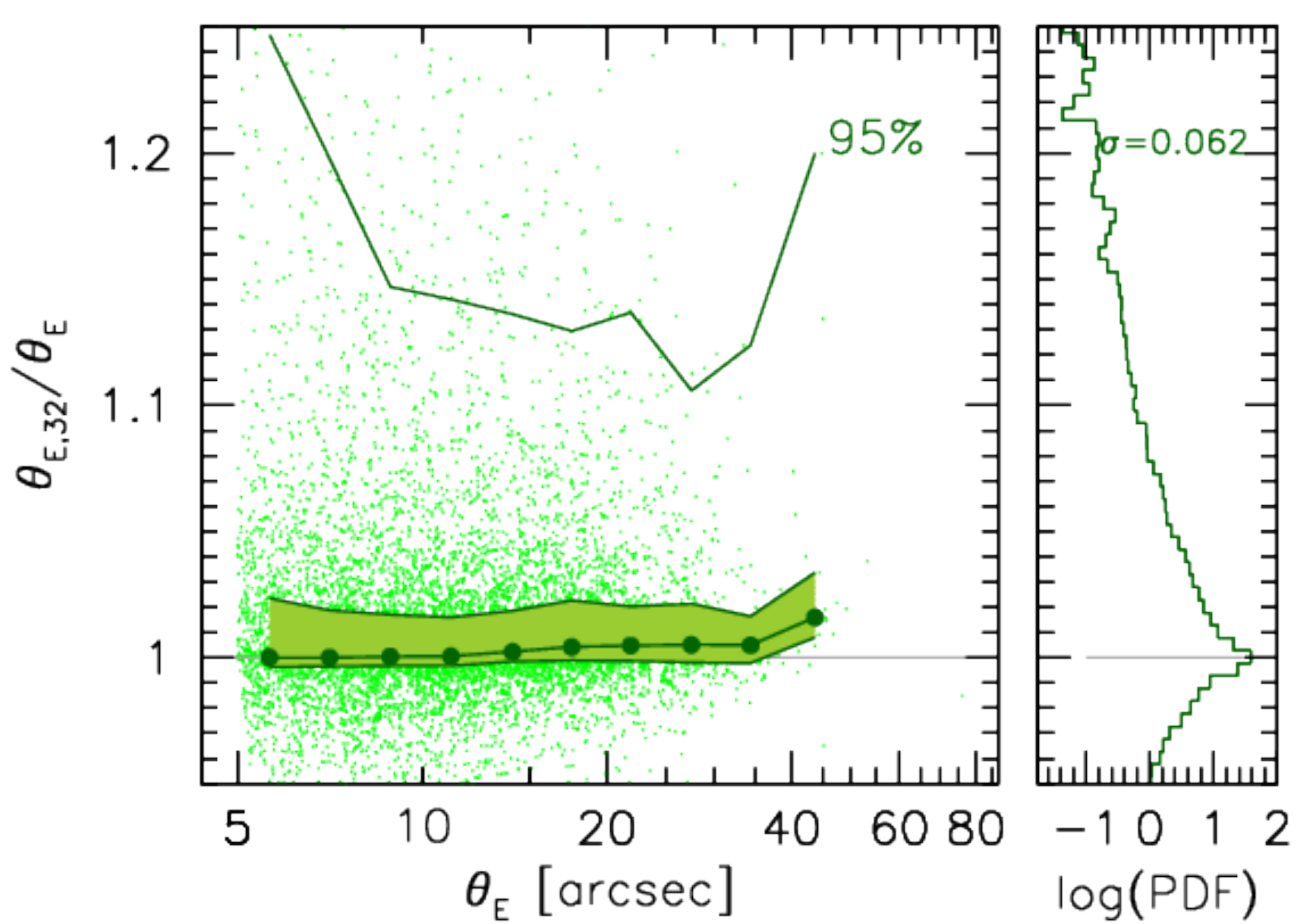}
\caption{Relative  change of  the  Einstein radius  of strong  lensing
  clusters extracted from the M-XXL simulation when we include all the
  matter from  a region  up to  $16$ Mpc/$h$  (left) and  $32$ Mpc/$h$
  along the  line-of-sight. As  a reference  we consider  the Einstein
  radii as computed from the run where  we select all the matter up to
  $8$  Mpc/$h$.  The  filled  circle  points  show  the  median  of  the
  distribution, while the  shaded area encloses the first  and the third
  quartiles of the distribution at fixed $\theta_E$.\label{fig16e32}}
\end{figure*}
However in order to give a hint on how much the Einstein radii change
including     more    matter     along    the     line-of-sight,    in
Figure~\ref{fig16e32} we show the relative size of the Einstein radius
-- with respect to the case in which we select a region of $8$ Mpc/$h$
along the line-of-sight -- computed selecting a region of $16$ Mpc/$h$
(left) and  $32$ Mpc/$h$  (right) along  the line-of-sight.   From the
figure we  can observe that  the median  value is consistent  with 
unity  (filled blue  points)  and  that, in  some  cases, large  scale
structures  may  boost  the  size  of  the  Einstein  radius  even  by
$30\%$, we notice also that  this population is less then $5\%$ 
of the whole sample. 
The  shaded regions in  the figure  enclose the first  and the
third  quartiles of  the  distribution at  fixed  $\theta_E$.  In  each
panel, the  solid line encloses $95\%$  of the data.  The  histogram in
the right sub-panels  shows the distribution along  the y-axis together
with the value of the standard deviation of the distributions.

\subsection{Strong lensing models of Clusters using the MOKA code}

Running  and analysing  large  numerical simulations  are usually  non
trivial tasks; in addition performing self consistent lensing analysis of
different cosmological  models including  also baryon  physics requires
large computational  resources and various  post-processing pipelines.
However, the results from  different numerical simulations -- $N$-body
only or including hydrodynamical processes -- can  be implemented using a pseudo-analytic
approach to construct convergence maps of different galaxy clusters   
in various physical  cases as, done with  the MOKA  code \citep{giocoli12a}.  
 MOKA pseudo-analytically
reconstructs high-resolution convergence  maps of galaxy  clusters
free from particles and numerical resolution limitations, implementing
results obtained from the most recent numerical runs.  
The virial mass of a halo is defined as
\begin{equation}
  M_{vir}         =          \frac{4         \pi}{3}         R_{vir}^3
  \frac{\Delta_{vir}}{\Omega_{m}(z)} \Omega_0 \rho_c\,,
\label{massdef}
\end{equation} 
where  $\rho_c$  represents  the  critical density  of  the  Universe,
$\Omega_0=\Omega_m(0)$  the matter  density parameter  at  the present
time     and    $\Delta_{vir}$     is    the     virial    overdensity
\citep{eke96,bryan98}, $R_{vir}$  symbolises the virial  radius of the
halo which  defines the distance  from the halo centre  that encloses
the desired density contrast. Haloes typically follow the NFW profile and 
are assumed to be triaxial -- following the model by \citet{jing02} and randomly oriented  with respect to the  line-of-sight. Each system is also populated by dark matter substructures assuming the subhalo population model by \citet{giocoli10a}.
In modelling the subhalo density profiles we account for tidal stripping due to close
interactions  with  the  main  halo  smooth  component  and  to  close
encounters  with other  clumps, gravitational  heating,  and dynamical
friction \citep{hayashi03,vandenbosch05,choi07,giocoli08b} 
using  a truncated   singular    isothermal   sphere \citep{metcalf01}.
For the halo concentration-mass relation we use the \citet{zhao09} model
 which  links  the   concentration of  a given halo  with the time ($t_{0.04}$)  at which
  its main progenitor assembles $4$ percent of its mass -- each halo mass 
  accretion history is computed using the results by \citet{giocoli12b}.
Haloes may also be  populated by galaxies
according to a HOD approach \citep{wang06} and once settled the central galaxy, with
a given stellar mass profile, the surrounding dark matter distribution
can  adiabatically   contract  \citep{blumenthal86,keeton01,gnedin11}.
To model the stellar mass density profile MOKA has two implementations: the
the   Hernquist \citep{hernquist90} and  Jaffe  \citep{jaffe83} profiles.
Both for the Hernquist and the Jaffe profiles we compute the 
central galaxy scale radius $r_g$ from the the half-mass (or effective)
radius $R_e$ by $r_g$ = 0.551$R_e$ and as done by \citet{keeton01}
we define the effective radius to be $R_e$ = $0.003 R_{vir}$. 
The contribution of all the components are then summed together
to compute the cluster convergence map as it can be read from the relation:
\begin{eqnarray}
\label{eqkappatot}
  \kappa(x,y) = \kappa_{DM}(x,y) &+&\\ 
  \kappa_{star}(x,y) &+& \sum_{i=1}^{N}\kappa_{sub,i}(x-x_{c,i},y-y_{c,i})\,,\nonumber 
\end{eqnarray}
where  $x_{c,i}$ and  $y_{c,i}$ represent  the coordinates of the center  of mass  of  the $i$-th
substructure.
The code is very  fast and allows not only to  study the dependence of
lensing observables on the halo properties but also to perform different
cosmological studies  \citep{boldrin12,boldrin16} and  comparisons with
various observational data \citep{xu15}, presenting great complementarity also with 
approaches used in various observational studies \citep{jullo07,more12}.

\begin{figure*}
\includegraphics[width=0.49\hsize]{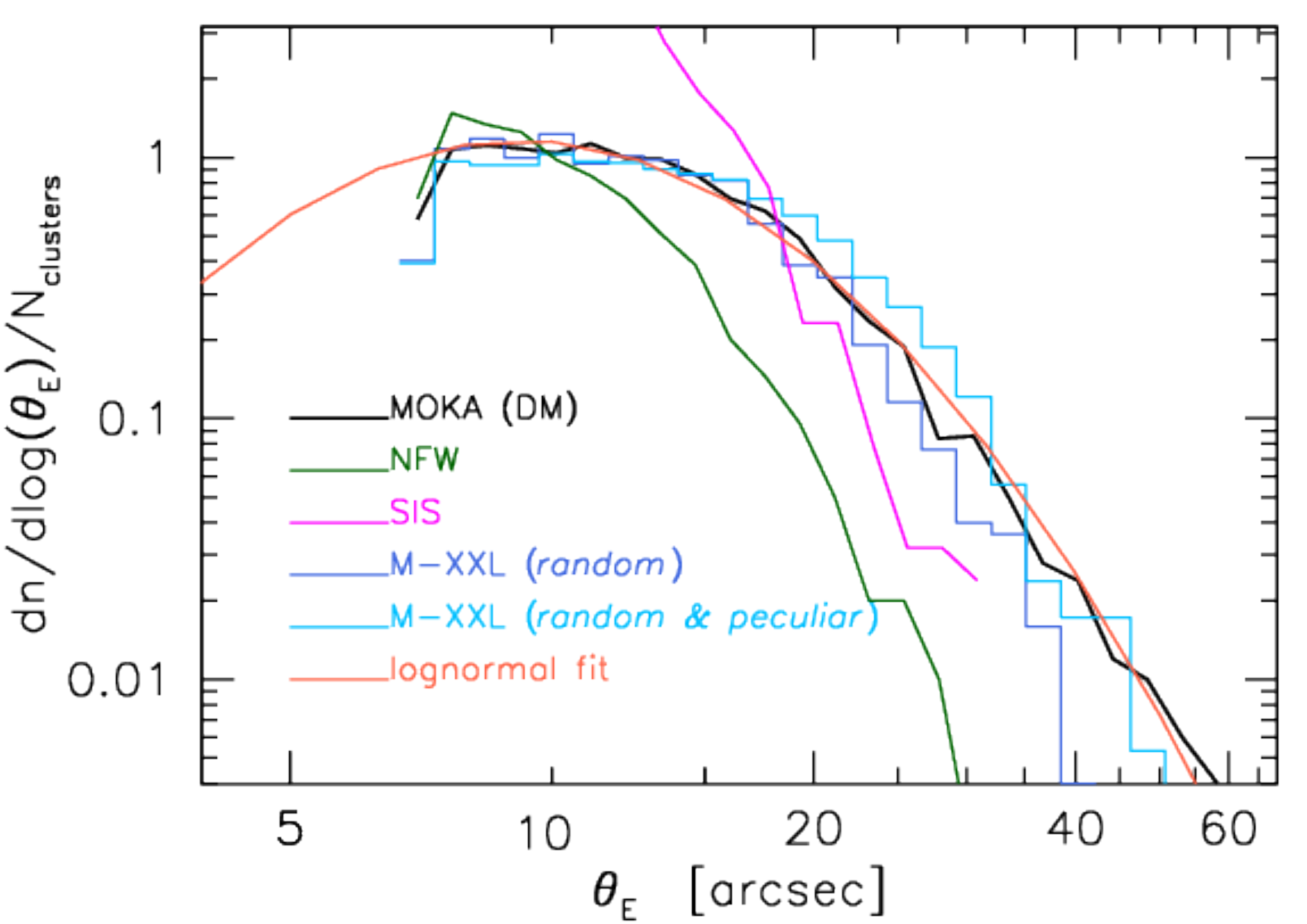}
\includegraphics[width=0.49\hsize]{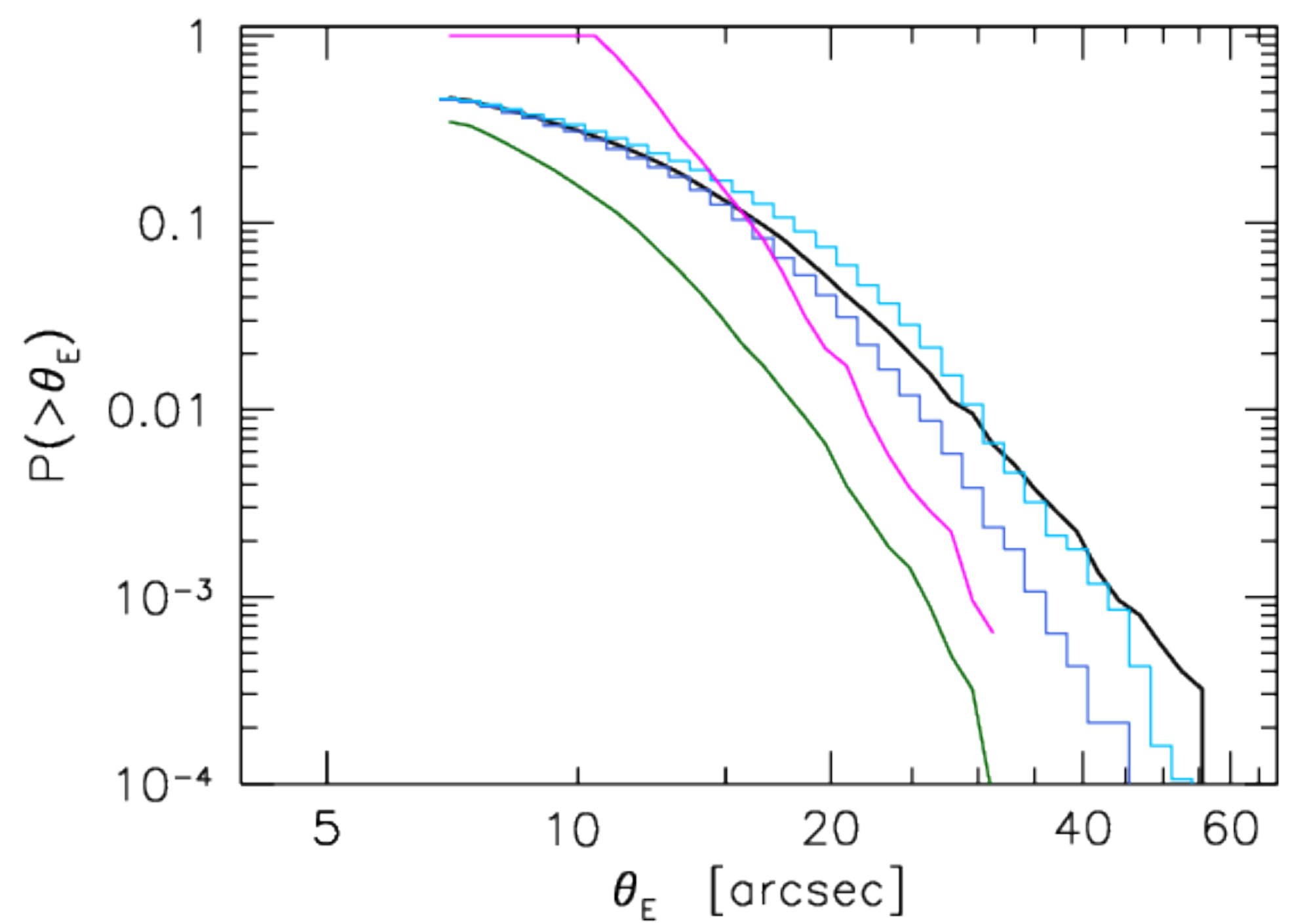}
\caption{Differential (left panel) and cumulative (right panel)  
Einstein radius distributions in
  M-XXL  and MOKA  clusters.  The  blue and  cyan histograms  show the
  distributions measured in the  M-XXL simulation considering only the
  \textit{random projections} and all the six ones,  respectively.  The
  black curve represents the predictions obtained  with a MOKA DM run on
  the  same cluster  masses,  while  the green  and  magenta ones  are the
  predictions for  the same masses assuming  a smooth NFW or  SIS halo
  profile for the lenses.  \label{figMOKAandMXXL}}
\end{figure*}

\begin{table*}
\centering
\caption{Summary of MOKA runs performed  with different models for the
  central galaxy.\label{tabruns}}
\begin{tabular}{r|c|c|c|c}
run  & triaxiality  & minimum  $m_{sb}$ &  BCG profile  & DM  Adiabatic
Contraction       \\        \hline       sDM       &        NO       &
$10^{10}M_{\odot}/h$\;\&\;$10^{12}M_{\odot}/h$ & NO & NO \\ DM & YES &
$10^{10}M_{\odot}/h$ &  NO & NO  \\ H  & YES &  $10^{10}M_{\odot}/h$ &
Hernquist &  YES \\  J  & YES & $10^{10}M_{\odot}/h$  & Jaffe &
YES \\ \hline
\end{tabular}
\end{table*}

In  Figure   \ref{figMOKAandMXXL}  we  compare  the   Einstein  radius
distributions  of  different MOKA  cluster  realisations  of the  mass
sample extracted from the M-XXL  catalogue at $z=1$ having $M_{200}>3
\times 10^{14}M_{\odot}/h$ and  $\theta_E>7$ arcsec. 
We remind the reader that for this comparison we did not
generate haloes from the corresponding theoretical mass function but we 
provide MOKA a sample of clusters with the same masses as the ones in the M-XXL
simulation extracted at $z=1$. Lenses
are then located at $z_l=0.5$ and sources at $z_s=2.5$. On the left we
show the differential distributions normalised  to the total number of
clusters while  on the  right the cumulative  ones.  To  be consistent
with the numerical simulation, MOKA haloes have been generated without
a Bright Central Galaxy  (BCG) and have a c-M relation  as for the M-XXL
cosmology at  $z=1$, computed using the  \citet{zhao09} model adopting
the \citet{giocoli12b}  halo mass  accretion history model.   The blue
and  the  cyan  histograms  show   the  measurements  from  the  M-XXL
considering only the three \textit{random projections} and considering
all   the   six   ones  --   \textit{random}   plus   \textit{peculiar
  projections}, respectively.  The MOKA maps  have been created with a
resolution of $1024\times 1024$ pixels  and are extended up the virial
radius scale of  the cluster. As in the M-XXL  analysis we compute the
lensing potential  and the  shear going  in the  Fourier space  and to
avoid  artificial boundary  effects  we  enclose the  maps  in a  zero-padded  region of  $512$  pixels.   The black  solid  line shows  the
predictions for different  MOKA realisations of the  same cluster mass
sample.   For comparison  the green  and the  magenta curves  show the
Einstein  radius distributions  computed assuming  a smooth  spherical NFW 
\citep{navarro96} and  Singular Isothermal  Sphere (SIS)  halo sample.
The latter profile is typically used to predict the location of strong
lensing images  in elliptical  galaxies \citep{koopmans06,koopmans09}.
We  estimate the velocity  dispersion of the
SIS halo we adopt the definition depending on the halo virial mass and
radius     according    to     the     spherical    collapse     model
\citep{wu98,cooray02}.  The figure suggests a quite good consistence in
the Einstein  radius statistics between  MOKA and M-XXL  clusters, and
also that a simple spherical NFW model highly under-predicts the strong
lensing  capability   of  SLCs   with  respect   to  a   triaxial  and
substructured  case.  As  done by  \citet{zitrin11c}, we  describe our
results with a log-normal distribution (red curve in the left panel of
the  figure):  the  relation  has $\mu  =  2.219$  and  $\sigma=0.532$
normalised consistently as done for the computed distribution from our
data; approximately $47\%$ of the  clusters with $M_{200} > 3 \times
10^{14}M_{\odot}/h$  possess  an  Einstein   radius  larger  than  $7$
arcsec.

\begin{figure}
\includegraphics[width=\hsize]{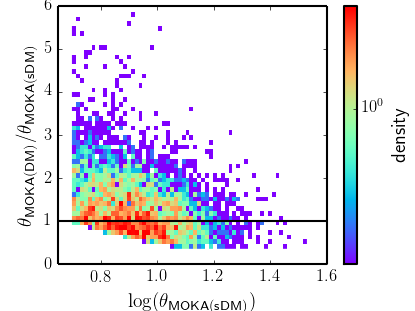}
\caption{Effect of triaxiality in distorting  the size of the Einstein
  radius.   Relative size  of  the Einstein  radius  between a  smooth
  spherical    and    triaxial    NFW   halo    as    obtained    from
  MOKA.\label{figtriax}}
\end{figure}

\begin{figure*}
\centering
\includegraphics[width=0.475\hsize]{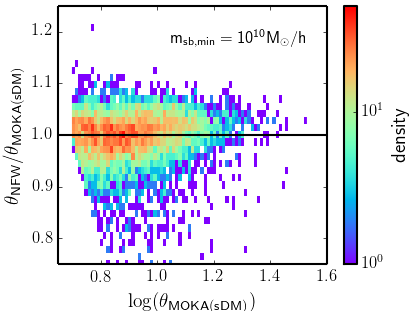}
\includegraphics[width=0.3\hsize]{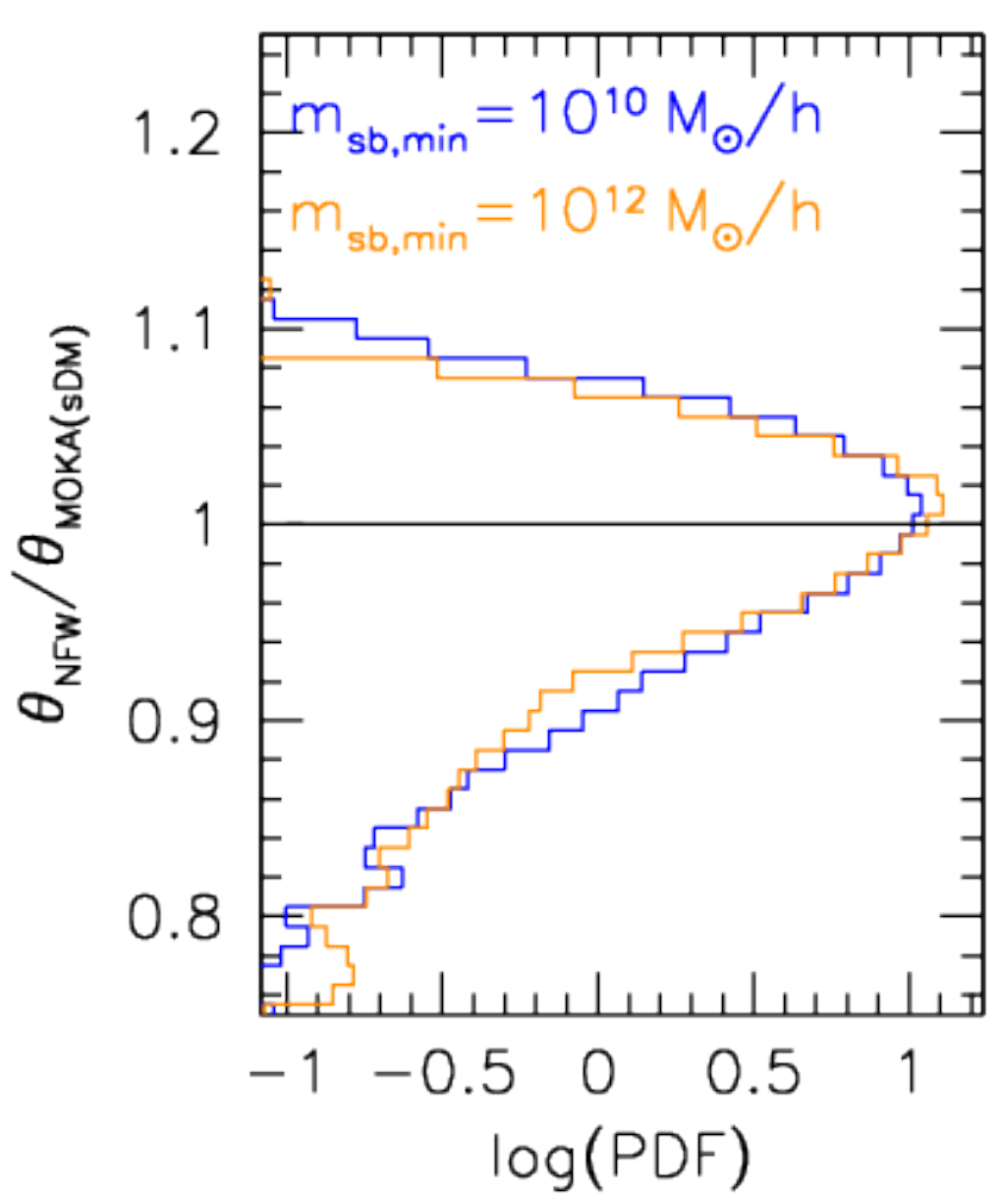}
\caption{\label{figsubs}Effect  of  substructures  in  perturbing  the
  Einstein radius of  galaxy clusters. Relative variation  of the size
  of the Einstein radius in  presence of substructures. To isolate the
  effect we  consider the case of  spherical clusters. In the  left panel we
  show the density probability  distribution of the relative variation
  assuming  $10^{10}M_{\odot}/h$  as  minimum subhalo  mass.  
  In  the right  panel we
  present the  Probability Distribution Function distributions  of the
  Einstein radius variation for the two subhalo minimum mass cases: $10^{10}$ and $10^{12}M_{\odot}/h$.}
\end{figure*}

In Figure~\ref{figtriax}  and \ref{figsubs}  we discuss the  effect of
triaxiality   and  substructures   on   the  size   of  the   Einstein
radius. Going step-by-step, in the first figure we compare the size of
the Einstein radii of the same sample of haloes when running MOKA with
the triaxiality  off (spherical  sDM)   and on (triaxial DM),
keeping identical all the other cluster and map properties. The effect
of  triaxiality, as  already discussed  by different  authors studying
haloes  in numerical  simulations  \citep{jing02,despali14}, is  quite
crucial and  typically tends  to boost  the Einstein  radius even  by a
factor of  four.  Nevertheless,  it is important  to observe  that the
scatter around  unity is quite  asymmetric and depends on  how the
halo ellipsoid -- typically prolate -- is oriented with respect to the
line-of-sight.  On the other  side, in Figure~\ref{figsubs} we isolate
the effect of substructures performing  two sets of simulations for our
halo sample.  In  the first we consider a spherical  DM halo populated
with substructures down to $m_{sb}=10^{10}M_{\odot}/h$ and in the second down
to $10^{12}M_{\odot}/h$ for the minimum subhalo mass and compare their
results with  respect to a  smooth and  spherical NFW sample.   In the
left panel  of the  figure we  compare the size  of Einstein  radii of
clusters in  our first run  with respect to  the size computed  from a
smooth NFW halo  with the same mass and concentration.   We remind the
reader that in  populating a halo with substructures we  use the model
by \citet{giocoli10a} for the subhalo mass function and the results by
\citet{gao04}  for the  subhalo distribution  in the  host.  Once  the
subhalo  mass  function  is  sampled  we compute  the  total  mass  in
subhaloes and subtract it to the input halo mass to compute the smooth
halo component, to which we assign  a concentration such that the mass
density profile of the smooth  plus clump components matches the input
assigned concentration.  In the right panel of Figure \ref{figsubs} we
show the  probability distribution  function of the  relative Einstein
radius variation  between the  smooth NFW  halo and  the substructured
runs  with the  two  different minimum  subhalo  mass thresholds.   We
notice that the  presence of small substructures tends  to perturb the
size of  the Einstein radius,  but  are the most  massive ones
that  mainly  contribute to  distort  the  strong lensing  regions  --
although this depends  on the relative distance of  the perturber from
the critical curves of the cluster.

A correct treatment of the mass density distribution in the central 
region of the cluster
is  very  important  for  strong lensing  modelings  and  predictions
\citep{meneghetti03}.    Numerical  simulations   and  semi-analytical
models          forecast          that          merger          events
\citep{springel01b,delucia04,tormen04}  that  drive the  formation  of
dark matter haloes  along the cosmic time bring to  the formation of a
massive  and   bright  galaxy  at   the  centre  of   galaxy  clusters
\citep{merritt85}. These central galaxies  are typically the brightest
galaxies in clusters and are usually  referred as Brightest Central Galaxy
(BCG).   They  are the  most  massive  galaxies  in the  Universe  and
generally   are   giant   ellipticals:   their   position   correspond
approximately to the  geometric and kinematic centre  of the cluster
and  to   the  position   of  the  peak   of  the   X-ray  emission.
Nevertheless, it  is interesting  to mention  that there  are clusters
where  these  conditions are  not  all  satisfied  at the  same  time:
typically this happens in systems  that are not completely relaxed and
present  merging  events \citep{katayama03,sanderson09,zhang16}.   For
the density distribution  of the stars in the BCG,  in our analyses we
make   use   of   two  different   parametrisations:   the   Hernquist
\citep{hernquist90} and the Jaffe  \citep{jaffe83} profiles.  We remind
the reader  that in  running MOKA with  these parameterisations  we ($i$)
assign  the   stellar  mass  to   the  BCG  using  a HOD  formalism
\citep{wang06},  ($ii$) conserve  the total  mass in  the cluster  and
($iii$) allow  the dark  matter density distribution  to adiabatically
contract \citep{keeton03,giocoli12a}.
\begin{figure*}
\includegraphics[width=0.49\hsize]{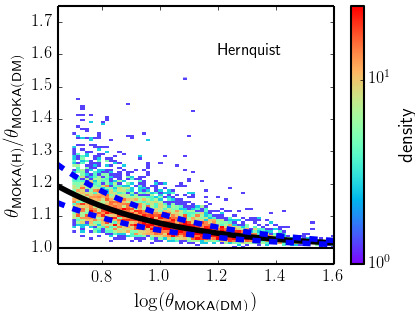}
\includegraphics[width=0.49\hsize]{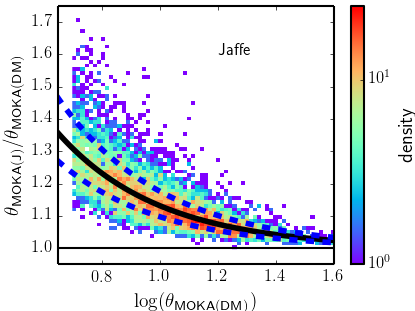}
\caption{Influence of the presence of a Bright Central Galaxy (BCG) in
  modifying  the size  of the  Einstein radius.  Left and  right panels
  refer to  the case  of a  BCG with a  Hernquist and  Jaffe profile,
  respectively.  The solid curve in  each panel represents the best fit
  relation  to   the  scatter  distribution   as  it can  be   read  from
  eq. (\ref{eqBCG}). The dashed  curve shows 1$\sigma$ uncertainties on
  the measured best fit parameters. \label{figBCG1}}
\end{figure*}
In  Figure \ref{figBCG1}  we show  the relative  size of  the Einstein
radius for strong lensing clusters (with $\theta_E> 5$ arcsec -- below
which our  measurements may be affected  by the grid size  of the map)
between  a  pure DM  run  and  a set  of  simulations  that assume  a
Hernquist (left panel) and a Jaffe profile (right panel) for  the BCG.  Because of
the  different behaviour  of the  profile, the  two models  give quite
different  relative  results. We  remind  the  reader that  while  the
Hernquist profile has  a logarithmic slope in the inner  part of $-1$,
the Jaffe profile goes like $-2$  and both profiles in the outskirts of
the BCG proceed like $-4$. From  the figure we notice that for smaller
values of   Einstein  radii the  results are  dominated by  the mass
density distribution of  the BCG, on the contrary for  larger values of
$\theta_E$ they are influenced by the dark matter profile: triaxiality
plus  clumpiness  of  the  halo.   In the  figures  the  solid  curves
represent the best fit relation to the scattered data points; they can
be, respectively, read as:
\begin{eqnarray}
\label{eqBCG} \theta_E/\theta_{E,(DM)} = &A& \exp\left(\beta \theta_{E,(DM)}\right)  \\
\mathrm{Hernquist} \;&A&=0.024^{+0.251}_{-0.281}\;\;\;\beta=-2.579^{+0.070}_{-0.051} \nonumber \\
\mathrm{Jaffe} \;&A&=0.817^{+0.217}_{-0.242}\;\;\;\beta=-2.843^{+0.075}_{-0.059}; \nonumber
\end{eqnarray}
the dashed curves show the $1\sigma$ uncertainties on the relations.

\begin{figure}
\includegraphics[width=\hsize]{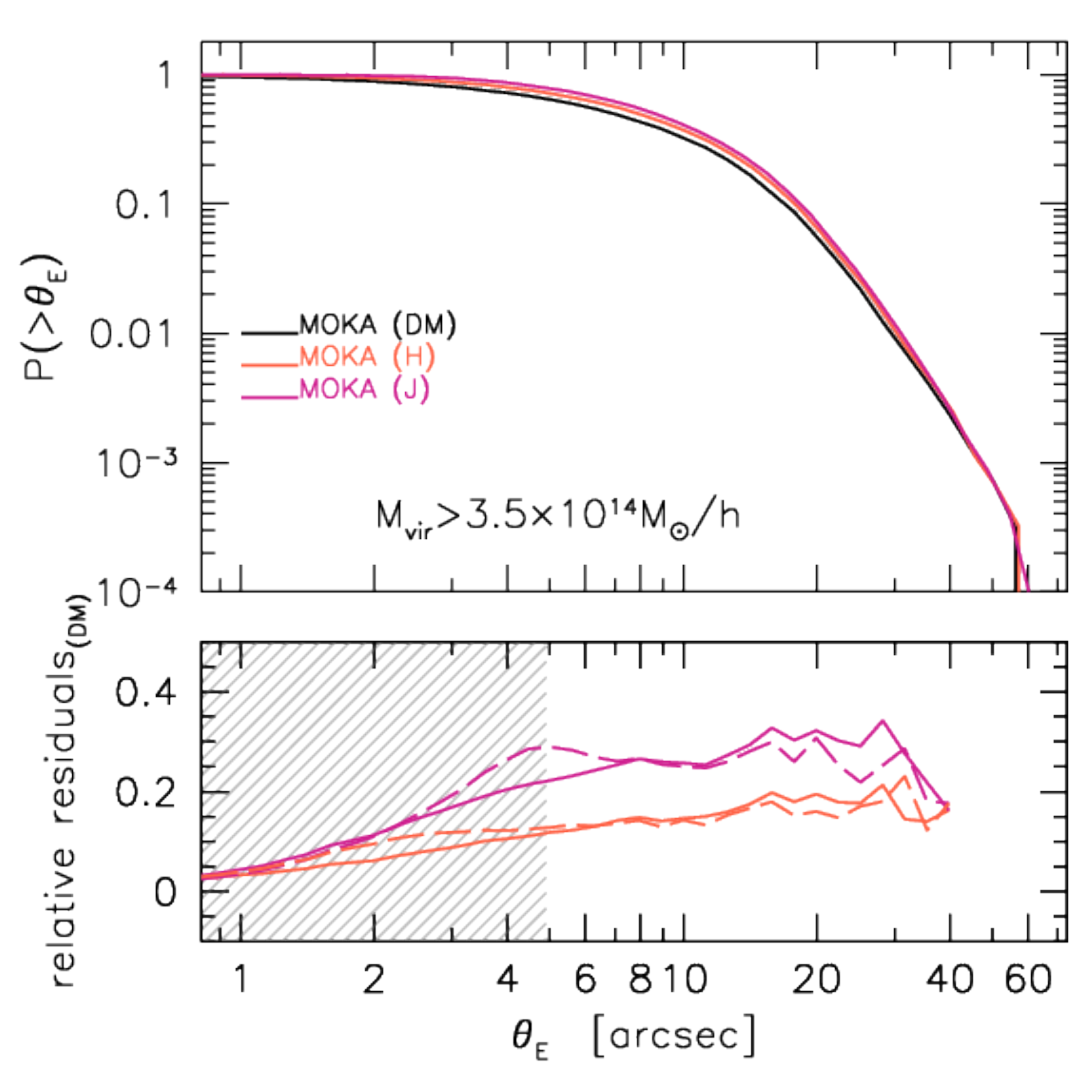}
\caption{Cumulative probability distribution of  having a cluster with
  an Einstein  radius larger  than a given  value.  Black,  orange and
  salmon coloured  curves show  the distribution  computed from  a MOKA
  Monte Carlo Simulation of the M-XXL mass sample assuming no BCG (DM)
  a BCG with a  Hernquist (H) or a Jaffe (J)  profile, respectively. The solid curves
  in the bottom panel show the relative residuals of the distributions
  with  respect to  the  DM only  case.  The  dashed  curves refer to  the
  predictions of the Einstein radius  distributions computed for the two
  cases   from    the   DM    run   using    the   relation    as   in
  eq. ~\ref{eqBCG}.\label{figBCG2cum}}
\end{figure}
The benefit of having these  fitting functions is illustrated in Figure
\ref{figBCG2cum},   where  we   compare  the   cumulative  probability
distribution computed from a pure DM run (black curve) and from the two runs
that assume the Hernquist (orange curve) and Jaffe (purple curve)  profiles for the
central galaxy. In the bottom panel, we present the relative residuals
of the cumulative distributions of the  two runs with BCG with respect
to the DM-only case. The presence of a  BCG increases the probability
of having a cluster with  an Einstein radius larger than $\theta_E>10$
arcsec  of about  $10-20\%$ --  depending on  the density  profile
model  of   the  central  galaxy.    In  the  bottom  panel   the  two
corresponding dashed  coloured curves show the  relative distributions
obtained  from  the  pure  DM  results  and  accounting  for  the  BCG
contribution sampling the best fitting relations as in eq.~\ref{eqBCG}
with their appropriate scatter for the parameters.

\section{Strong Lensing Scaling Relations}
\label{sslsr}

In this section we discuss the correlation of the size of the Einstein
radius with  other galaxy  cluster properties. In  the figures  we will
present we  have used  the only-DM MOKA  runs; the  relations obtained
considering runs with a BCG following a Hernquist  and a Jaffe profile will be
summarised in Table~\ref{tabscaling}.

\begin{figure}
\includegraphics[width=\hsize]{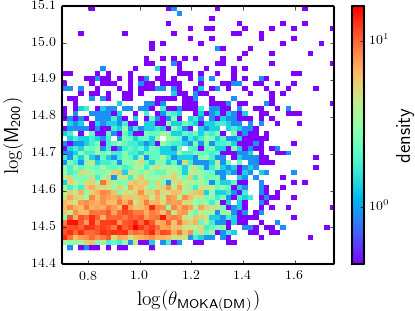}
\caption{Correlation between the cluster  masses and the measured size
  of the Einstein radii from the two-dimensional convergence maps. The
  colours show the point  density of clusters of the DM-only
  MOKA      run      in      the      $\log(M_{200})$-$\log(\theta_E)$
  space. \label{figM200thetaE}}
\end{figure}

The first correlation  we have considered is between  the cluster mass
and the size of the  Einstein radius (see Fig~\ref{figM200thetaE}).  In
this case, as  it is evident from the figure, the two  quantities do not
show a  good correlation: we argue  that this is probably due to
the fact that what mainly matters in shaping the Einstein radius is the
halo triaxiality, the concentration and the presence of substructures;
but we are observing in the plane of the sky a random orientation of
the cluster.   The second attempt  has been done by correlating  the weak
lensing  mass of  the  clusters  with the  Einstein  radii.  Using  a
Navarro-Frank-White (NFW,  \citep{navarro96,navarro97}) model  for the
matter density profile, it is possible to compute the associated shear
profile $\gamma_{NFW}$ once the lens and  the source redshifts
have been fixed.
For each simulated  convergence map we compute  the spherical average
shear profile  and measure the  associated weak lensing  mass $M_{wl}$
and concentration $c_{wl}$ by minimising
\begin{equation}
\chi^2(M_{wl},c_{wl}) = \dfrac{\sum_i \left[ \gamma(r_i) - \gamma_{NFW}(M_{wl},c_{wl})\right]^2}{\sigma_{g,i}^2};
\label{chi2wl}
\end{equation}
 $\sigma_{g,i}$ represents the shape measurement error computed assuming a
number  density  of  sources  of $20$/arcmin$^{2}$  --
mimicking the  number of  background sources expected  to be  usable for
weak lensing measurements in future space-based observations:
\begin{equation}
  \sigma_{g,i} = \dfrac{\sigma_{\epsilon}}{\sqrt{n_g A_i}}\,
\end{equation}
being $A_i$ the area of the  $i$-annulus.  This method gives typically a
good determination of  the projected mass responsible  for the lensing
signal        \citep{hoekstra12,giocoli14,vonderlinden14a}.
To summarise for each cluster -- and for each of the corresponding run
(DM, Hernquist and  Jaffe) -- we have measured of  the associated weak
lensing mass  and concentration in  addition to the evaluation  of the
size of the Einstein radius.
\begin{figure}
\includegraphics[width=\hsize]{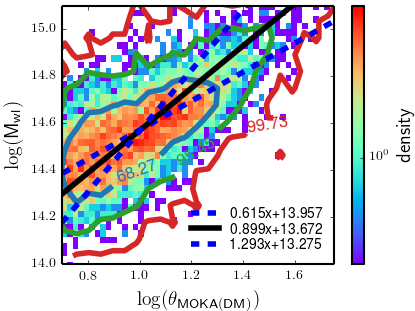}  
\caption{Scaling relation  between the weak lensing  estimated mass $M_{wl}$ --
  computing the  best fit profile with  a NFW functional to  the weak
  lensing shear profile -- and the size of the Einstein radius. In the
  figure we  show the  relation for  the DM-only  run; results  of the
  other  runs  are  summarised in  Table~\ref{tabscaling}.   The  blue
  dashed lines show  the best least-squares fit to  the data points in
  the       $\log(M_{wl})-\log(\theta_E)$       and       in       the
  $\log(\theta_E)-\log(M_{wl})$   spaces,   while   the   black   line
  represents the bisector  of them. The blue, green  and red contours
  enclose $68.27$, $95.45$ and $99.73\%$
  of the data points, respectively.
\label{scalingMwl}}
\end{figure}
In  Figure~\ref{scalingMwl}  we present  the  scaling  relation --  in
logarithmic space and for the DM-only  run -- between the weak lensing
cluster  mass $M_{wl}$  and  the corresponding  size  of the  Einstein
radius.   The blue  dashed lines  show the  least-squares  fits in  the
$\log(M_{wl})-\log(\theta_E)$ and in the $\log(\theta_E)-\log(M_{wl})$
spaces, while the  black line indicates the  corresponding bisector of
them.  The  blue, green  and red  curves enclose  $68.27$, $95.45$ and $99.73\%$  of  the data  points, respectively.   The third  relation ---
which  offers a  very strict  correspondence --  we have  considered is
between the  size of the  Einstein radius associated with  a NFW-halo
having $M_{wl}$  and $c_{wl}$ and  the one  we measure from  our maps.
For  a NFW  halo we  compute $\theta_E(M,c)$  -- given  its mass  and
concentration in addition to the lens and source redshifts -- from the
profile of $[1-\kappa(\theta)-\gamma(\theta)]$ and measuring the angular
scale   where  the   relation  $1-\kappa(\theta_E)-\gamma(\theta_E)=0$
holds.
\begin{figure} 
\includegraphics[width=\hsize]{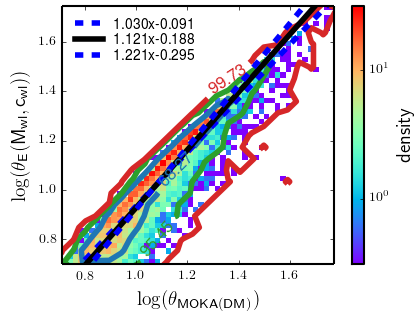}
\caption{Scaling relation between  the Einstein radius of  a NFW halo
  with the same mass and concentration -- computed by fitting with a NFW
  functional the  weak lensing shear  profile --  and the size  of the
  Einstein radius of the cluster.  As in Fig.~\ref{scalingMwl} we show
  the   correlation   only   for   the   DM   run   and   present   in
  Table~\ref{tabscaling} the  results for  the runs with  BCGs. Dashed
  blue  lines  show   the  least-squares  fit  to  the   data  in  the
  $\log(\theta_E(M_{wl},c_{wl}))-\log(\theta_E)$     and     in     the
  $\log(\theta_E)-\log(\theta_E(M_{wl},c_{wl}))$   spaces,  while   the
  black line  displays the bisector of  them. \label{scalingEt}}
\end{figure}
In  Figure~\ref{scalingEt} we  display  this correlation  -- again  in
$\log_{10}$ space for  the DM-only run:
the results for the runs with BCGs are reported in 
 Table~\ref{tabscaling}. As  in the
previous  figure, the  coloured contours  enclose  enclose  $68.27$, $95.45$ and $99.73\%$  of  the data  points, the dashed blue  lines displays least-squares
fit  in the  $\log(\theta_E(M_{wl},c_{wl}))-\log(\theta_E)$ and  in the
$\log(\theta_E)-\log(\theta_E(M_{wl},c_{wl}))$ spaces,  while the black
line their  bisector. In  this case we notice  that the correlation between  the two
quantities is very  close, apart for some asymmetries  -- some systems
possess an Einstein radius larger  than the one computed from $M_{wl}$
and $c_{wl}$. A case by case  analysis of these systems has brought to
two main causes: ($i$) large value  of the ellipticity in the plane of
the sky and ($ii$) projection of  substructures close to the centre of
the cluster. Both these give at the end a poor NFW fit to the shear profile,
leading to  an  underestimation  of  the  corresponding  Einstein  radius
associated to  the NFW profile. We found that the first  cause is
the most probable in the majority of the cases.

\begin{table}
\centering
\caption{Best fit  linear coefficients --  $y=m x +  q$ -- of  all the
  runs  (DM,  Hernquist  and  Jaffe)  of  the  relation  displayed  in
  Fig.~\ref{scalingMwl}  and  \ref{scalingEt}.   The  superscript  and
  underscript numbers present the coefficient  of the least squares in
  the $x-y$ and $y-x$ planes, respectively.
  \label{tabscaling}}
\begin{tabular}{r|c|c}
  run & $m$ & $q$ \\\hline
  & $\log(M_{wl})-\log(\theta_E)$ & \\\hline\hline
  DM &  $0.899^{0.615}_{1.293}$ & $13.672^{13.957}_{13.275}$   \\\hline
  Hernquist &  $0.892^{0.652}_{1.206}$ & $13.641^{13.884}_{13.322}$ \\\hline
  Jaffe &  $0.894^{0.694}_{1.143}$ & $13.589^{13.791}_{13.336}$  \\ \hline\hline
  & $\log(\theta_E(M_{wl},c_{wl}))-\log(\theta_E)$ & \\\hline \hline
  DM &  $1.121^{1.030}_{1.221}$ & $-0.188^{-0.091}_{-0.295}$  \\\hline
  Hernquist & $1.153^{1.066}_{1.249}$ & $-0.226^{-0.132}_{-0.330}$   \\\hline
  Jaffe &  $1.168^{1.089}_{1.254}$ & $-0.238^{-0.153}_{-0.331}$ \\ \hline  \hline
\end{tabular}
\end{table}

\section{Sensitivity of the Einstein radius distribution on $\Omega_M$ and $\sigma_8$}
\label{scerd}

The  number density  and the  properties of  galaxy clusters  has been
investigated  by different  authors  with the  aim  of understanding  how
sensible      they      are       to      cosmological      parameters
\citep{rozo10,waizmann12,waizmann14,boldrin16,sartoris15}.   In this  section
we discuss  how the  Einstein radius  distributions are  sensitive 
to  the total  matter density in  the Universe  $\Omega_M$ ---
assuming   always  to   live   in  a   flat   universe:  $\Omega_M   +
\Omega_{\Lambda}=1$  --  and  to the   linear  mass  density  r.m.s. on a scale of $8$ Mpc/$h$ $\sigma_8$ -- that defines
the initial  matter power spectrum normalisation.   We always consider
lenses located at $z_{l} = 0.5$ and to obtain their abundance we
randomly sample the \citet{sheth99b} mass  function between $z = 0.48$
and $z=0.52$.
\begin{figure}
\includegraphics[width=\hsize]{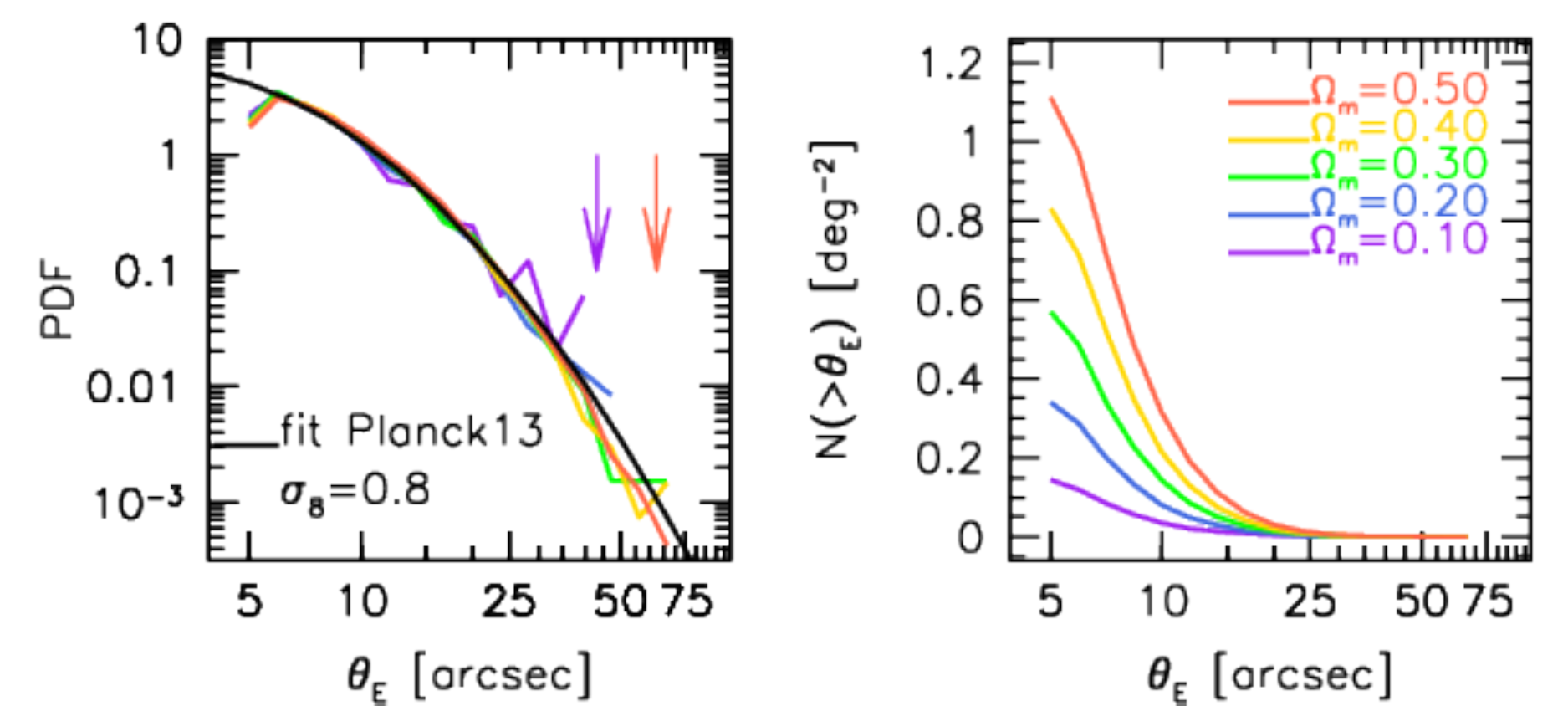}
\includegraphics[width=\hsize]{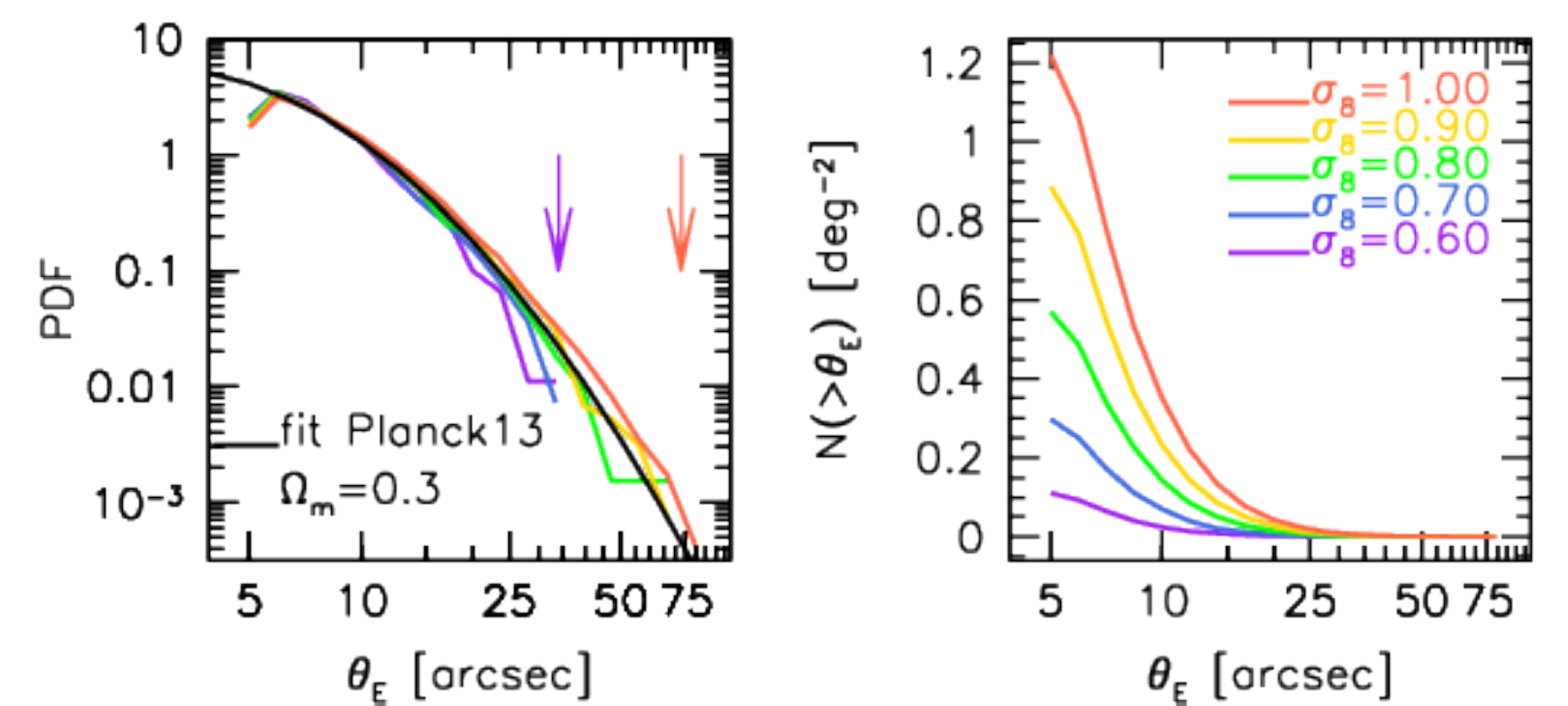}
\caption{Left  panels:  probability   distribution  functions  of  the
  Einstein radius distributions of a Monte Carlo realisation of lenses
  at  redshift $z_l=0.5$  with  sources located  at  $z_s=2.5$ --  the
  cluster number  density has been computed  from the \citet{sheth99b}
  mass  function integrated  on  the whole  sky  between $z=0.48$  and
  $z=0.52$. Right  panels: cumulative  number counts of  strong lenses
  per  square degrees  with an  Einstein  radius larger  than a  fixed
  value. Top and bottom panels display the case of varying $\Omega_M$ and
  $\sigma_8$ at  a time,  respectively.  The black  curve in  the left
  panels represents  the log-normal  relation (eq.~\ref{eqfitPlanck13})
  that   better  describes   the   Planck13  probability   distribution
  function. The  arrows on the  left panels mark the  largest Einstein
  radius find in the two extreme corresponding models.
  \label{thetaEcosmos1}}
\end{figure}
In  Figure~\ref{thetaEcosmos1} we  show  the Probability  Distribution
Function  (left  panels)  and  cumulative number  density  per  square
degrees (right  panels) of  the Einstein  radii in  cosmological model
with different $\Omega_M$ (top  panels) and $\sigma_8$ (bottom panels)
parameters -- fixing  one at a time.  The vertical  arrows -- coloured
according  to the  corresponding  cosmological model  -- indicate  the
largest Einstein  radius found in  the two extreme samples  assuming a
full sky  realisation.  From  the figure we  notice that  the Einstein
radii  regularly  increase with  $\Omega_M$  and  $\sigma_8$; this  is
because galaxy  clusters are  more numerous  in these  cosmologies but
also because  they are also  more concentrated; in addition  we remind
that clusters  at $z_l=0.5$  in universes  with higher  matter content
possess also a  lower lensing distance $D_{lens}$.   We underline that
the BCG treatment is absent in these simulations and remind the reader
that  the counts  can be  adapted to  the two  considered BCG  stellar
density  profiles   at  the   light  of   the  results   discussed  in
Figures~\ref{figBCG1} and  \ref{figBCG2cum}.  In  the left  panels the
black  solid  curves  display  the log-normal  best  relation  to  the
Planck13 counts that can be read as:
\begin{equation}
 \mathrm{PDF}(\theta_E) = \dfrac{1}{\sqrt{2 \pi \sigma^2}} \exp{\left[-\dfrac{\left(\ln(\theta_E)-\mu\right)^2}{2 \sigma^2}\right]},
\label{eqfitPlanck13}
 \end{equation}
with $\mu=1.016$ and  $\sigma = 0.754$.  From the right  panels of the
figure we notice that a change of $\Omega_M$ -- or $\sigma_8$ -- of
$10\%$  corresponds  approximately to   a  variation  in  the  number   of  lenses  with
$\theta_E>5$ arcsec of about $20\%$.

\begin{figure}
\includegraphics[width=\hsize]{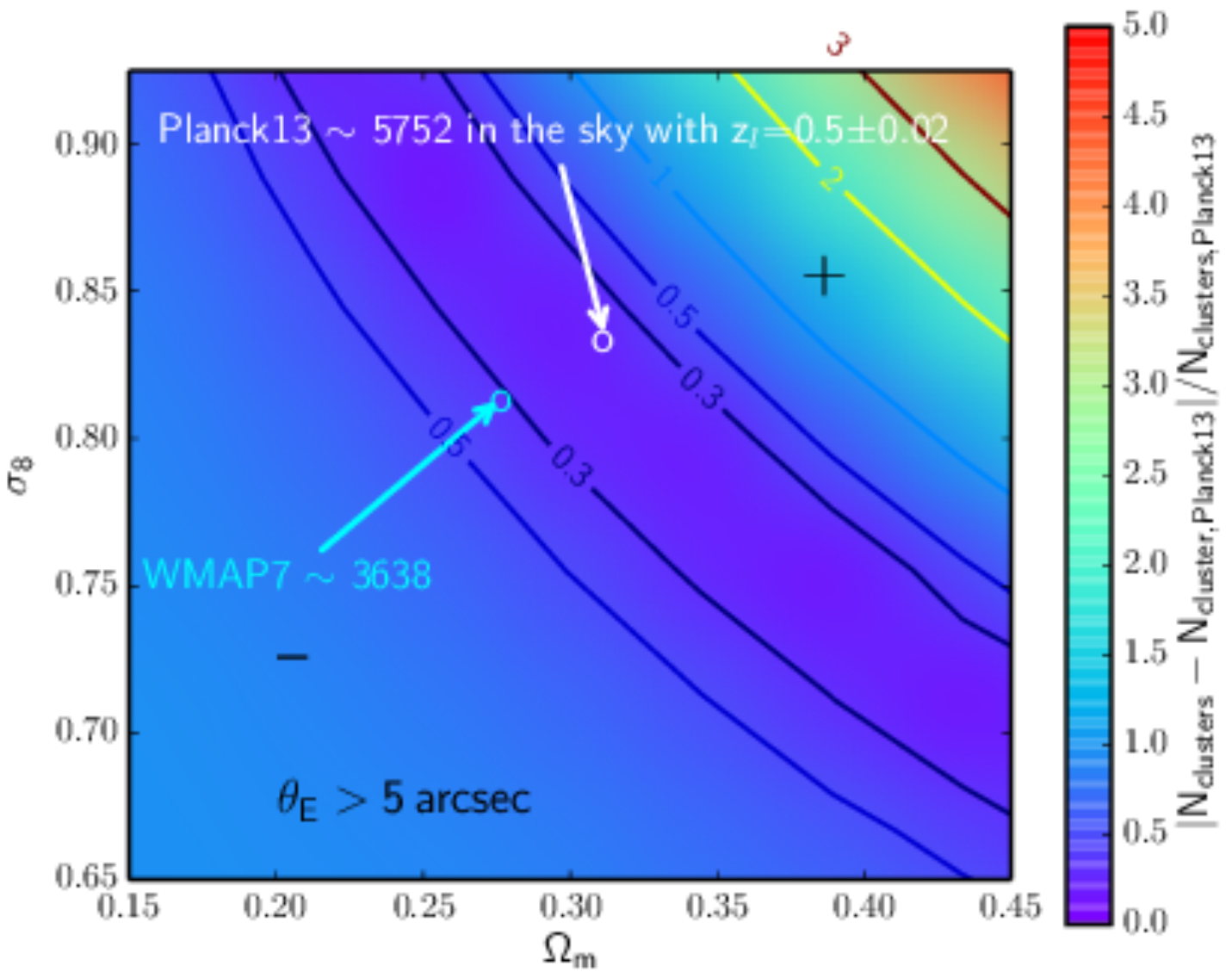}  
\caption{Relative counts of clusters  with Einstein radius larger then
  $5$  arcsec,   with  respect   to  the   Planck13  counts,   in  the
  $\Omega_M-\sigma_8$ plane.  The black  and the blue circles indicate
  the  counts on  the full  sky for  a Planck13  and WMAP7  cosmology,
  respectively.   We consider  the  number density  of cluster  lenses
  between $z=0.48$ and $z=0.52$, with sources at $z_s=2.5$. A plus and
  minus  signs in  the plane  indicate  the regions  where counts  are
  positive  and  negative  with  respect  to  the  Planck13  cosmology,
  respectively.\label{thetaEcosmos2}}
\end{figure}

In  Figure  \ref{thetaEcosmos2}  we  display the  relative  counts  of
clusters with $\theta_E$  larger then $5$ arcsec, with  respect to the
number computed for a Planck13 cosmology, in the $\Omega_M - \sigma_8$
plane.  As reference the black and the blue circles indicate the predicted
numbers of these strong lensing clusters in the  whole sky  between  $z_l=0.48$ and  $z_l=0.52$ for  the
Planck13  (black  circle)  and  WMAP7 (blue  cross)  cosmologies.  
For the WMAP7 model we assume 
$\Omega_M =0.272$, $\Omega_{\Lambda} = 0.728$, $h = 0.704$ and $\sigma_8 = 0.809$.
These results show that the expected counts  of the Einstein radii in these
two  models may differ  by more  than $3 \sigma$  assuming a
Poisson distribution: $5752$ for the Planck13 and $3638$ for the WMAP7 cosmology, respectively. 
We remind the reader that in computing the cluster counts for the various 
cosmological models we have accounted for the change of volume between $z=0.48$ and $z=0.52$. In particular the WMAP7 model has a volume (in $\left(\mathrm{Mpc}/h\right)^3$) of about $5\%$ larger 
than the Planck13 one because of a higher Hubble constant and lower total matter content, but fewer 
strong lensing cluster counts. This highlights that the change of volume is quite negligible 
with respect to the role played by  the initial power spectrum normalisation 
parameter $\sigma_8$ and by the different total matter content $\Omega_M$ for the SLC counts: in Planck13 we 
find more clusters and  those are more concentrated then in the WMAP7 cosmology because formed at higher 
redshifts. 
 As already noticed by \citet{boldrin16}  the degeneracy
relation  of the  SLC counts  behaves as  the cluster  counts plus the
 evolution of the  halo structural properties in different
cosmological models and the variation of the lensing distance.

\section{Summary and Conclusions}
\label{ssummary}
In this paper we have  presented the strong lensing properties of
a  sample  of  galaxy   clusters  extracted  from  the  Millennium-XXL
simulation analysing the distribution  of their Einstein radius. The results
have been  compared with a  Monte Carlo  MOKA realisation of  the same
mass sample  finding very good  agreement.  We have also  performed an
analysis   to  understand   how  sensitive   is  the   Einstein  radius
distribution on specific cosmological  parameters creating a sample of
clusters in  different models using  the MOKA  code. We find  that the
Einstein  radius distribution  is  quite sensitive  to $\Omega_M$  and
$\sigma_8$, as it  is the cluster abundance, and that  universes with high
values of $\Omega_M$  and $\sigma_8$ possess a large  number of strong
lensing clusters.

In the following points we summarise the main results of our analyses:
\begin{itemize}
\item a large  fraction of strong lensing  clusters are systematically
  biased by projection effects;
\item the orientation matters: when the major axis of the cluster ellipsoid is oriented along 
the line-of-sight the Einstein radius may be boosted by more then a factor of two with respect to a random orientation;
\item  the shape  of the  strong lensing  population is  slightly more
  triaxial than the overall considered cluster sample;
\item  a  self-consistent treatment  of  the  effects of  large  scale
  structures is  important for strong lensing  predictions: correlated
  systems may  boost the  Einstein radius of  galaxy clusters  by even
  more than $30\%$;
\item the  comparison between M-XXL  clusters and MOKA  realisations on
  the same sample  of cluster masses shows consistent  results for the
  Einstein  radius distribution,  and is well  described by  a log-normal
  distribution;
\item a  correct treatment of  the subhalo population and  the cluster
  triaxiality is important for  an adequate strong lensing modelling:
  typically the triaxiality matters more and  may boost the size of the
  cluster Einstein radii by various orders of magnitudes;
\item the  presence of  a Bright  Central Galaxy  in a cluster  tends to
  modify the total projected mass profile and consistently the size of
  the Einstein  radius; we have  discussed and modelled the  impact of
  two      different      stellar      mass      density      profiles
  \citep{hernquist90,jaffe83}  on the  strong  lensing properties  and
  noticed  that  the  Einstein  radii may  change  by  $10$-$60\%$  --
  depending on the cluster properties and stellar profiles;
\item the size of the Einstein radius has a quite tight correlation with
  the Einstein radius of a NFW  halo where mass and concentration are
  computed performing  a fit  to the weak lensing  shear profile;
  less stringent is the correlation with the weak lensing mass;
\item the Einstein radius distribution is very sensitive to the matter
  content  of  the Universe  ($\Omega_M$)  and  to the  initial  power
  spectrum normalisation  ($\sigma_8$):  we have noticed that
  universes with  larger values of  those parameters possess  a higher
  number of strong lensing clusters; this can help
  distinguish Planck13 and WMAP7 cosmology at $3$ $\sigma$.
\end{itemize} 

In conclusion,  our results encourage  in shading more light  into the
dark components  of our Universe  through the study of  strong lensing
cluster  populations,  foreseeing  the  unique results  that  will  be
available from the next-generation wide field surveys from space.

\section*{Acknowledgments}
CG  thanks CNES  for financial  support.  ML  acknowledges the  Centre
National de  la Recherche Scientifique  (CNRS) for its  support.  This
work was performed using facilities offered by CeSAM (Centre de donneS
Astrophysique de Marseille- (http://lam.oamp.fr/cesam/). This work was
granted  access  to  the  HPC resources  of  Aix-Marseille  Universite
financed by the project  Equip@Meso (ANR-10-EQPX-29-01) of the program
''Investissements d'Avenir``  supervised by the Agence  Nationale pour
la Recherche  (ANR).  This work  was carried  out with support  of the
OCEVU Labex (ANR- 11-LABX-0060)  and the A*MIDEX project (ANR-11-IDEX-
0001-02) funded by the  ``Investissements d'Avenir'' French government
program managed by the ANR.  We acknowledge support from the Programme
National  de Cosmologie  et Galaxie  (PNCG).  MM  acknowledges support
from  Ministry  of  Foreign  Affairs  and  International  Cooperation,
Directorate  General for  Country Promotion,  from INAF  via PRIN-INAF
2014 1.05.01.94.02, and from ASI via contract ASI/INAF/I/023/12/0.  LM
acknowledges  the grants  ASI n.I/023/12/0  "Attivit\`a relative  alla
fase  B2/C per  la missione  Euclid",  MIUR PRIN  2010-2011 "The  dark
Universe and the cosmic evolution  of baryons: from current surveys to
Euclid"  and PRIN  INAF  2012  "The Universe  in  the box:  multiscale
simulations  of  cosmic  structure".  REA  acknowledges  support  from
AYA2015-66211-C2-2.   EJ acknowledge  CNES support.   CG thanks  Mauro
Sereno,  Jesus Vega  and Michele  Boldrin for  useful discussions  and
python  stratagems.   We  also  thank  Giuseppe  Tormen  and  Vincenzo
Mezzalira  for  giving  us  the  possibility  to  use  their  computer
facilities  on which  part  of  the MOKA  simulations  have been  run.
\bibliographystyle{mn2e}

\bsp	

\bibliography{../../../../../AstroBib/globalbibs}
\label{lastpage}
\end{document}